\documentclass[]{iopart}

\usepackage{graphicx}
\usepackage{dcolumn}
\usepackage{bm}

\begin{document}

\title{Nonlinear dynamics in an alternating gradient guide for neutral particles}

\author{M. R. Tarbutt and E. A. Hinds}

\address{Centre for Cold Matter, Blackett Laboratory, Imperial
College London, SW7 2AZ, United Kingdom}
\ead{m.tarbutt@imperial.ac.uk}

\date{\today}

\begin{abstract}
Neutral particles can be guided and focussed using electric field gradients that focus in one transverse direction and defocus in the other, alternating between the two directions. Such a guide is suitable for transporting particles that are attracted to strong electric fields, which cannot be guided using static fields. Particles are only transmitted if their initial positions and transverse speeds lie within the guide's phase space acceptance. Nonlinear forces are always present in the guide and can severely reduce this acceptance. We consider the effects of the two most important nonlinear forces, a term in the force that is cubic in the off-axis displacement, and a nonlinear term which couples together the two transverse motions. We use approximate analytical techniques, along with numerical methods, to calculate the influence of these nonlinear forces on the particle trajectories and on the phase space acceptance. The cubic term alters the focussing and defocussing powers, leading either to an increase or a decrease of the acceptance depending on its sign. We find an approximate analytical result for the phase space acceptance including this cubic term. Using a perturbation method we show how the coupling term leads to slow changes in the amplitudes of the transverse oscillations. This term reduces the acceptance when it reduces the focussing power, but has little influence when it increases that power. It is not possible to eliminate both nonlinear terms, but one can be made small at the expense of the other. We show how to choose the guide parameters so that the acceptance is optimized.
\end{abstract}


\maketitle

\section{Introduction}

A neutral particle in an electric field gradient feels a force. The field polarizes the particle and, if it has a gradient, pulls on the induced dipole moment. In its ground state, a neutral particle is always attracted to regions of strong electric field. In other quantum states, the attraction may either be to strong fields, or to weak fields. Here, we are concerned with the guiding and focussing of strong-field-seeking particles. Static electric fields cannot do the job and so a dynamic focussing scheme needs to be used \cite{Auerbach66, Ketterle92}. An alternating gradient (AG) focusser is a series of lenses, each focussing in one transverse direction and defocussing in the other. The focus and defocus directions alternate. Particles have stable trajectories through the focusser because they are close to the axis when they travel through the defocussing lenses, and further away when passing through the focussing ones.

The purpose of an alternating gradient guide is to transmit a large number of particles over arbitrarily long distances with as little loss as possible. Such a guide might employ an electrode structure that is uniform along the beamline, with the alternation achieved by switching the applied voltages - alternation in time. Alternatively, the voltages can be fixed with the alternating gradients set up by the electrode structure itself - alternation in space. Early experiments with alternating gradient structures used two-rod and four-rod geometries to focus polar molecules, e.g. \cite{Kakati71,Lubbert78}. These same ideas were later applied to cold atoms, demonstrating that an atomic beam or fountain could be focussed this way \cite{Noh00, Kalnins05}. Of considerable interest at present is the application of AG focussing to the Stark deceleration of strong-field seeking molecules. Prototype alternating gradient decelerators have been demonstrated experimentally \cite{Bethlem02, Tarbutt04, Wohlfart08}, and their focussing properties investigated in detail both theoretically and experimentally \cite{Bethlem(1)06}. For heavy molecules to be decelerated to rest, these decelerators will need to employ very many stages of AG focussing, typically 100 lenses or more. Recently, a long AG guide with a double bend has been used to guide slow molecules from an effusive source \cite{Junglen04}. As buffer gas cooling technology advances, it is likely that such guides will be more widely used as efficient tools for extracting molecules from the cold buffer gas \cite{Maxwell05, Patterson07}. An AG guide has also been used to separate conformers of C$_6$H$_7$NO by exploiting the selectivity of the guide to the ratio of dipole moment to mass \cite{Filsinger08}. The theoretical analysis we present here is suitable for describing all of these linear guiding structures. It can also be applied directly to other alternating gradient structures such as a storage ring for strong-field seeking molecules \cite{Nishimura04} or linear ac traps of the kind demonstrated for both molecules \cite{Schnell07} and atoms \cite{Kishimoto06}. The physics we present also carries over directly to ac traps with cylindrical symmetry because the same terms appear in the equations of motion \cite{Bethlem(2)06}.

The anharmonic forces present in AG guides and traps for neutral particles tend to be detrimental to the phase space acceptance, often reducing this by a factor of 10 or more. Similarly, these nonlinearities reduce the quality of the images in an AG focussing system. This problem has long been recognized. In reference \cite{Lubbert78} for example, molecules were guided in a four rod geometry with the rod positions carefully chosen to optimize the linearity of the forces, resulting in a large improvement in transmission over previous two rod geometries. The optimization of electrode geometry with respect to minimizing nonlinear forces has been discussed in \cite{Kalnins02} and \cite{Bethlem(1)06}. Numerical simulations have been used to predict the performance of several types of AG guides \cite{Junglen04}, decelerators \cite{Bethlem(1)06} and traps \cite{Bethlem(2)06, Schnell07} all confirming the hugely deleterious effects of the nonlinearities.

In this paper we use a mixture of analytical and numerical techniques to calculate the effect of the nonlinear forces on the particle trajectories and the phase space acceptance of an AG guide. We start by reviewing the linear theory since this is the foundation for the analysis of nonlinearities. It is well known that the motion consists of an oscillation driven at the frequency of alternation, called the `micromotion', superposed on a slower oscillation of larger amplitude called the `macromotion'. A good approximate description of the macromotion is obtained using an effective potential. The anharmonic terms modify this effective potential and so alter the phase space acceptance. The leading anharmonic terms in this potential are a quartic term along with a term that couples together motion in the two transverse directions. The quartic term in the potential leads to terms in the force that are cubic in the transverse coordinates. We treat this quartic term analytically, deriving an expression for the phase-space acceptance in the presence of this term. To understand the effect of the transverse coupling term we apply a perturbation method which reveals the nature of the coupled trajectories and shows that this term tends to decrease the phase space acceptance even when it increases the depth of the effective potential. We use numerical techniques throughout to support the theory, test the approximations and obtain results when the perturbation approach ceases to be valid. We apply our analysis to some example AG guides for neutrals.

\section{Equations of motion}

Our starting point is equation (8) of reference \cite{Bethlem(1)06}, which gives a suitable form for the electrostatic potential inside an AG guide, $\Phi(x,y)$, as a function of the transverse coordinates $x$ and $y$:

\begin{equation}
\fl \Phi(x,y) =  \Phi_{0}\left(a_{1}\frac{x}{r_{0}} +
a_{3}\frac{\left(x^{3} - 3xy^{2}\right)}{3r^{3}_{0}} + a_{5}\frac{\left(x^{5} - 10x^{3}y^{2} +
5xy^{4}\right)}{5r^{5}_{0}} \right).
\label{Eq:phi}
\end{equation}

\noindent This equation is derived by writing down a multipole expansion of the potential, and then eliminating terms that are unsuitable for a guide. The scale factors, $r_0$ and $\Phi_{0}$ characterize the size of the electrode structure and the applied voltages, while $a_{1}$, $a_{3}$ and $a_{5}$ are dimensionless constants representing the sizes of the dipole, hexapole and decapole terms in the expansion.

A neutral particle inside the guide will be polarized by the electric field to a degree that depends on the particle polarizability and on the magnitude of the field. The interaction of this induced polarization with the electric field changes the potential energy of the particle; this is the Stark shift, $W$. If the electric field is spatially inhomogeneous, so too is $W$, and there is an associated force acting on the particle, ${\bf F}=-{\bf \nabla} W$. For our present purpose, we do not need to know the detailed dependence of the Stark shift on $|E|$, the magnitude of the electric field. We need only know that, for sufficiently small electric fields, the Stark shift is quadratic in $|E|$, $W=-\frac{1}{2} \alpha |E|^{2}$, while for sufficiently large electric fields it becomes linear, $W = - \mu |E|$. The constants of proportionality in the two cases are known as the polarizabilty, $\alpha$, and the dipole moment, $\mu$. Whether the electric field is considered small or large depends on the polarizability of the particle. For example, the Stark shift of a CaF molecule is in the linear regime at a field of 100\,kV/cm, whereas the Stark shift of a Li atom is quadratic in the same field.

We now impose the additional constraint that $a_{5} \ll a_{3} \ll a_{1}$, meaning that the electric field in the guide is dominated by a constant term, and that the decapole term is much smaller than the hexapole term. We then obtain the following approximate expression for the $n$'th power of the electric field magnitude, valid throughout the region where $x,y < r_{0}$:

\begin{eqnarray}
\fl |E(X,Y)|^{n} =  E_{0}^{n} ( 1 + a_{3}n(X^{2} - Y^{2}) &+ (a_{3}^{2}n(3-n) - 6 a_{5}n) X^{2}Y^{2} \nonumber
\\* &+ (a_{3}^{2} n (n-1)/2 + a_{5} n) (X^{4} + Y^{4}) + \cdots).
\label{Eq:E}
\end{eqnarray}

\noindent In this equation, we have set the redundant scale parameter $a_{1}$ to unity and have introduced scaled transverse coordinates $X=x/r_{0}$, $Y=y/r_{0}$. $E_{0}$ is the electric field at the centre of the guide. The neglected terms are of order $a_{3}a_{5}$, $a_{3}^{3}$ and higher. Since $n$ can be any number, we see from (\ref{Eq:E}) that the potential in which the particles move always has the same functional dependence on the transverse coordinates, irrespective of whether the Stark shift is linear, quadratic, or
intermediate. Although we have treated the case of a long guide, similar terms appear in the treatment of cylindrically symmetric traps \cite{Bethlem(2)06}, and so our analysis could also be readily applied to that case.

In writing the equations of motion, we may either use the axial coordinate, $z$, or the time, $t$, as the independent variable. The former is most appropriate for long guiding structures, while the latter tends to be used for trapping geometries. We will use $z$ throughout, writing the equations of motion as

\begin{eqnarray}
X''(Z) = \eta^{2}\left(\frac{a_3}{|a_3|}X + \chi Y^{2}X + \zeta X^{3}\right), \nonumber \\
Y''(Z) = \eta^{2}\left(-\frac{a_3}{|a_3|}Y + \chi X^{2}Y + \zeta
Y^{3}\right), \label{Eq:EqnOfMotion}
\end{eqnarray}

\noindent where the primes denote differentiation with respect to
the scaled axial coordinate $Z = z/r_{0}$, and $\eta$, $\chi$ and
$\zeta$ are dimensionless parameters that depend on the character
of the Stark shift, but not on the spatial coordinates. For a
linear Stark shift, $\chi = 2|a_{3}|-6 a_{5}/|a_{3}|$, $\zeta = 2
a_{5}/|a_{3}|$, and

\begin{equation}
\eta^{2}={\cal R}|a_{3}|=\frac{\mu E_{0} |a_{3}|}{1/2M v_{z}^{2}}.
\label{Eq:etaSqLin}
\end{equation}

\noindent For a quadratic Stark shift, $\chi = |a_{3}|-6
a_{5}/|a_{3}|$, $\zeta = |a_{3}|+ 2 a_{5}/|a_{3}|$, and

\begin{equation}
\eta^{2}={\cal R}|a_{3}|=\frac{\alpha E_{0}^{2} |a_{3}|}{1/2M
v_{z}^{2}}. \label{Eq:etaSqQuad}
\end{equation}

\noindent We have introduced the quantity $\cal R$, which
characterizes the Stark shift on the axis of the guide in terms of
the forward kinetic energy, and will be useful later in the paper.
We have written the equations of motion for a lens that focusses
in the $y$-direction when $a_{3}$ is positive. For focussing in
the $x$-direction, simply interchange $X$ and $Y$.

We note from (\ref{Eq:EqnOfMotion}) that the dominant term in
the force is linear in the transverse coordinate, focussing in one
plane and defocussing in the orthogonal plane. In addition to the
desirable linear term, there is a term cubic in the transverse
displacement, and also a nonlinear transverse coupling term. By a
suitable choice of the parameters $a_{5}$ and $a_{3}$, it is
possible to eliminate one or other of the nonlinear terms, but
simultaneous elimination of both nonlinear terms is impossible
without also setting $\eta$ to zero. For example, when the Stark
shift is linear, the cubic term in the force can (in principle) be
eliminated by choosing $a_{5}=0$; alternatively, the coupling term
can be removed by choosing $a_{5}=a_{3}^{2}/3$; the elimination of
both terms is clearly not possible. Thus, the lenses in an
alternating gradient guide for neutral particles are necessarily
aberrant. More generally, the values of $\zeta$ and $\chi$ are
linearly related via $\chi + 3\zeta = 2n|a_3|$. This relationship
shows that it is also impossible for both $\chi$ and $\zeta$ to be
negative, implying, as we will show later, that no matter how the
electrodes are arranged at least one of the terms is detrimental
to the transmission of the guide.

In (\ref{Eq:EqnOfMotion}), we are considering the case where $a_5 \ll a_3 \ll 1$, and so the coefficient of the linear term is much larger than the coefficients of the nonlinear terms. It is therefore tempting to suppose that the effect of the nonlinearity on the trajectories is small. However, in going from a focussing lens to a defocussing lens, the linear term changes sign (as it must) but the nonlinear terms do not. The alternating gradient guide can support a stable beam because the envelope of the beam is always smaller in the defocussing lenses than in the focussing lenses. For this reason, the net effect of the linear forces is to confine the molecules, but this trajectory-averaged confining force is considerably weaker than the linear forces of the individual lenses. By contrast, the nonlinear forces are identical in the two lenses and simply add up. It follows that, even though the nonlinear terms in the individual lenses are small, their role is crucial in understanding the dynamics of the guided particles.

\section{Motion in the linear approximation}

Throughout this paper we consider a linear AG guide consisting of
alternate converging and diverging lenses each of length $L$,
separated by gaps of length $S$. To understand the effect of the
nonlinear forces, we first analyze the motion in the absence of
the nonlinearities. This idealized motion has been discussed in
some detail in reference \cite{Bethlem(1)06} and is based on the
original work of Courant and Snyder \cite{Courant58} who discussed
the same problem in the context of charged particle focussing. We
give a brief summary of this theory here, and extend the
discussion of reference \cite{Bethlem(1)06}.

\subsection{Trajectories}
Neglecting the nonlinear terms, we write the equation of motion in
one dimension as

\begin{equation}
X''(Z) + K(Z)X(Z) = 0. \label{Eq:EqMotionLinear}
\end{equation}

\noindent Here, $K(Z) = \eta^{2} G(Z)$, where G(Z) is a function
whose value is $+1$ in a focussing lens, $-1$ in a defocussing
lens, and 0 in a drift space. We know the solution in each region
of constant $G$. For example, the solution inside a focussing lens
is given by

\begin{eqnarray}
\left( \begin{array}{cc} X \\ X' \end{array} \right)_{Z_{0}+Z}  &=
\left(
\begin{array}{cc} \cos(\eta Z)
& \eta^{-1} \sin(\eta Z) \\
-\eta \sin(\eta Z) & \cos(\eta Z)
\end{array} \right)\left( \begin{array}{cc} X \\ X'
\end{array} \right)_{Z_{0}} \nonumber\\
&= F(\eta, Z)\left( \begin{array}{cc} X \\ X'
\end{array} \right)_{Z_{0}} \label{Eq:FocusMatrix}
\end{eqnarray}

\noindent The solution inside a defocussing lens is identical to (\ref{Eq:FocusMatrix}), with the replacement of $\eta$ by $i \eta$, and the transfer matrix denoted by $D(\eta,Z)$. For a
region of free space, let $\eta \Rightarrow 0$ in (\ref{Eq:FocusMatrix}) and call the matrix $O(Z)$.

Knowing the solutions in each of the separate regions, we can
construct the full trajectories piecewise, but the Courant-Snyder
formalism offers a far better approach. This involves looking for
a general solution of the form

\begin{eqnarray}
X(Z) &= \sqrt{\epsilon \beta(Z)}\cos(\psi(Z)+\delta) \nonumber \\
&= A_{1}\sqrt{\beta(Z)}\cos \psi(Z) + A_{2}\sqrt{\beta(Z)}\sin
\psi(Z) \label{Eq:trajEq}
\end{eqnarray}

\noindent where $\beta$ is a $Z$-dependent amplitude function that
has the same periodicity as the AG array, $\psi$ is a Z-dependent
phase, and $\epsilon$, $\delta$, $A_{1}$ and $A_{2}$ are defined
by the initial conditions.

By substitution we find that (\ref{Eq:trajEq}) is a valid
solution to (\ref{Eq:EqMotionLinear}) provided

\begin{equation}
\psi(Z) = \int_{0}^{Z} \frac{1}{\beta(\tilde Z)}\,d\tilde Z
\label{Eq:psi}
\end{equation}

\noindent and

\begin{equation}
-\frac{1}{4} \beta'^{2} + \frac{1}{2}\beta \beta'' + K \beta^{2} =
1. \label{Eq:diffBeta}
\end{equation}

\noindent It appears that all we have achieved is the replacement of the
differential equation for $X(Z)$ by a more complicated
differential equation for $\beta(Z)$. The advantage is that we
will never have to solve this new differential equation. We
already know the solution to (\ref{Eq:EqMotionLinear}), but
piecewise, and our aim is to express this in a convenient form.
From (\ref{Eq:trajEq}) we have

\begin{equation}
X'(Z) = \frac{A_{1}}{\sqrt{\beta}}(-\alpha \cos \psi - \sin \psi)
+ \frac{A_{2}}{\sqrt{\beta}} (-\alpha \sin \psi + \cos \psi)
\label{Eq:xprime}
\end{equation}

\noindent where

\begin{equation}
\alpha = -\frac{1}{2} \beta'. \label{Eq:alpha}
\end{equation}

\noindent Using (\ref{Eq:trajEq}) and (\ref{Eq:xprime}) we find the
relationship between the coordinates $X,X'$ at position $Z$ and
those at position $Z+L_{{\rm cell}}$, $L_{{\rm cell}}$ being the
periodicity of the AG array. We make use of the periodicity
constraint on $\beta$, $\beta(Z+L_{{\rm cell}}) = \beta(Z)$ (and
hence also on $\alpha$) and thus obtain:

\begin{equation}
\fl\left( \begin{array}{cc} X \\ X'
\end{array} \right)_{Z+L_{{\rm cell}}} = \left( \begin{array}{cc} \cos \mu + \alpha \sin\mu
& \beta \sin \mu \\
-\gamma \sin \mu & \cos \mu - \alpha \sin \mu \end{array} \right) \left( \begin{array}{cc} X \\ X'
\end{array} \right)_{Z}, \label{Eq:CSM}
\end{equation}

\noindent where $\gamma = (1 + \alpha^{2})/\beta$ and

\begin{equation}
\mu = \psi(L_{{\rm cell}}) = \int_{Z}^{Z+L_{{\rm cell}}}
\frac{1}{\beta(\tilde Z)}\,d\tilde Z \label{Eq:mu}
\end{equation}

\noindent is the phase advance per unit cell. Since the integral
is taken over a full period, $\mu$ is independent of $Z$. The
matrix appearing in (\ref{Eq:CSM}) is known as the
Courant-Snyder matrix, ${\cal M}$.

The procedure to obtain $\beta(Z)$ is now straightforward. We
illustrate this procedure for a simple guide with no gaps. Consider an
alternating sequence of focussing and defocussing lenses, each of
length $L$, with $Z=0$ defined to be at the entrance of a
focussing lens. We first obtain an explicit expression for the
transfer matrix between a point $Z$ inside a focussing lens and
the equivalent point one lattice unit further downstream. This
matrix is $M=F(\eta, L-Z).D(\eta, L).F(\eta, Z)$. We then
calculate $\case{1}{2}\Tr(M)$, which, from the Courant-Snyder matrix is
equal to $\cos \mu$. The explicit result is $\cos \mu = \cos(\eta
L)\,\cosh(\eta L)$. Finally, we calculate $\beta$ by equating the
upper right hand element of $M$ to that of ${\cal M}$, obtaining

\begin{equation}
\beta(Z) = \frac{\cosh(\eta L)\sin(\eta L) + \cos(\eta L - 2\eta
Z) \sinh(\eta L)}{\eta \sqrt{1-\cos^{2}(\eta L)\cosh^{2}(\eta
L)}}, \label{Eq:explicitBeta}
\end{equation}

\noindent for $0 \le Z \le L$. We can apply the same procedure to
a point $Z$ inside a defocussing lens (i.e. $L < Z < 2L$). The
result for $\beta(Z)$ is identical to (\ref{Eq:explicitBeta})
with the replacement of $\eta$ by $i \eta$ and the replacement of
$Z$ by $Z-L$. Since $\beta$ has period $2L$, we now know it
everywhere, and since $\psi(Z)$ can be calculated directly from
$\beta(Z)$ we have the complete general solution to the equation
of motion in the form of (\ref{Eq:trajEq}). This same
procedure can equally well be applied to more complicated guiding
structures.

\subsection{\label{Sec:LinearPSA}
Phase-space acceptance}

\begin{figure}
\centering
\includegraphics[width=9cm]{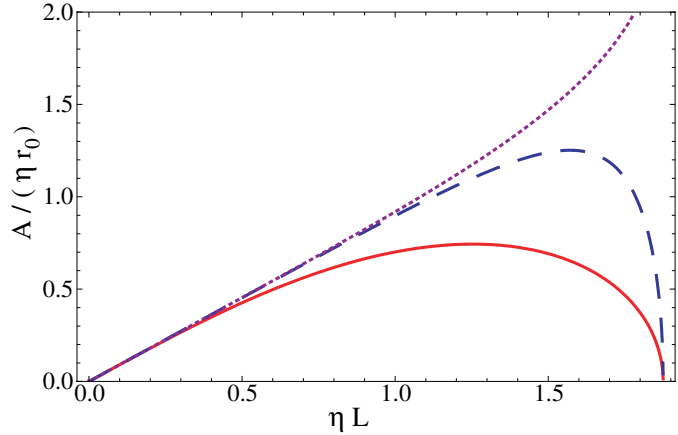}
\caption{1D phase-space acceptance, in units of $\eta r_{0}$, as a
function of the parameter $\eta L$. Solid line: Uniform aperture
throughout the guide. Dashed line: Aperture at defocussing lenses
only. Dotted line: Approximate result for small micromotion. \label{Fig:LinAcc}}\end{figure}

Returning to the case of a general guide, particles enter with a
range of transverse positions and angles and we wish to know which
of them are transmitted. Consider a particle whose initial
conditions are given by $\epsilon$ and $\delta$, as in (\ref{Eq:trajEq}). Using the first
line of (\ref{Eq:trajEq}) to form the quantity $X^{2} +
(\alpha X + \beta X')^{2}$, and recalling that $\beta' =
-2\alpha$, $\psi'=1/\beta$ and $1+\alpha^{2} = \gamma \beta$, we
find the invariant

\begin{equation}
\gamma X^2 + 2 \alpha X X' + \beta X'^2 = \epsilon.
\label{Eq:CSInvariant}
\end{equation}

Suppose we record the transverse position and angle of the
particle at the longitudinal positions $Z + N L_{cell}$, for a
large number of integer values of $N$. Then a plot of these points
on a phase-space diagram, whose axes are $X$ and $X'$, traces out
the ellipse defined by (\ref{Eq:CSInvariant}). If we repeat
this procedure for a different value of $Z$, the shape of the
ellipse evolves periodically with $Z$ according to (\ref{Eq:CSInvariant}), but its area remains the same, $\pi \epsilon$. A distribution of particles having $\epsilon_{i} \le
\epsilon$ fills the ellipse. If the guide has a constant
transverse aperture with walls at $X=\pm 1$, then it is clear from (\ref{Eq:trajEq}) that all particles having $\epsilon_{i} < \epsilon$ will be transmitted indefinitely provided that
$\sqrt{\beta_{max} \epsilon} < 1$, where $\beta_{max}$ is the
maximum value of $\beta(Z)$. If this condition is satisfied, the
transverse extent of the beam never exceeds the transverse
aperture of the guide. Particles that have $\epsilon_{i} >
1/\beta_{max}$ may also be transmitted by the guide provided that
the turning points of the macromotion (the cosine factor in (\ref{Eq:trajEq})) never coincide with the turning points of
the micromotion (the amplitude function in (\ref{Eq:trajEq})). That condition may occur for a large
number of particles if $\mu/\pi$ is a rational number, $\mu$ being
the phase advance per unit cell defined by (\ref{Eq:mu}).
However, these ``resonances'' in the phase-space acceptance become
very narrowly centred on these special values of $\mu$ once the
guide is long compared to the wavelength of the macromotion, and
we can usually ignore them. Then, the transverse phase-space
acceptance is simply $\pi/\beta_{max}$. As discussed in
\cite{Bethlem(1)06}, the maximum value of $\beta$ always occurs in
the centre of a focussing lens. Returning to our specific example
of a gapless guide, this is explicitly seen in (\ref{Eq:explicitBeta}) at $Z=L/2$. It follows that, in this case, the transverse phase space-space acceptance in one dimension is

\begin{equation}
A_{\rm 1D} = \frac{\pi \eta \sqrt{1 - \cos^{2}(\eta L)
\cosh^{2}(\eta L)}}{\cosh(\eta L)\sin(\eta L) + \sinh(\eta L)}.
\label{Eq:exampleAcceptance}
\end{equation}

\noindent Since we are using $Z$ as the independent variable, the
acceptance has the units of $X X'$, i.e. $r_{0}\cdot {\rm
radians}$. There are, of course, two transverse directions, and
the 2D transverse acceptance is simply the square of the above
formula.

So far, we have supposed that the transverse aperture formed by
the electrodes of the guide is uniform along its entire length.
For many electrode geometries, this is a poor approximation. An
interesting case is that of alternating two-rod lenses, discussed in
\cite{Bethlem(1)06} and used in all experimental work on
alternating gradient deceleration reported to date. Here, the
electrodes lie in the defocussing plane and there is no aperture
at all in the focussing plane. The maximum beam envelope still
occurs at the centre of the focussing lens, but there is no
aperture at this point for the beam to crash into. Instead, the
appropriate limiting value of $\beta$ to insert into the
acceptance equation is its value at the entrance to the
defocussing lens. As a result, the phase-space acceptance
increases. For the case of a gapless guide with an aperture at the defocussing lenses only, the second term in the denominator of (\ref{Eq:exampleAcceptance}) is replaced
by $\cos(\eta L)\sinh(\eta L)$. The acceptance in one transverse
direction of a gapless guide is plotted in figure \ref{Fig:LinAcc}, for both kinds of transverse aperture.
We see that when the aperture is defined only by the defocussing
lenses (dashed line) the 1D acceptance peaks at a slightly higher
value of $\eta L$ where it is almost twice as large. Later in this paper, we will make a ``small micromotion approximation'' which amounts to neglecting the modulation of $\beta$, replacing it by the constant value $L_{\rm cell}/\mu$. In this approximation, the phase space acceptance in one dimension becomes $\pi \mu/L_{\rm cell}$. For a gapless guide, this approximate acceptance is shown by the dotted line in figure \ref{Fig:LinAcc}. We see that the approximate result agrees well with the exact results when $\eta L \ll 1$, because the micromotion amplitude is indeed small in this regime. As $\eta L$ increases, the approximate result overestimates the acceptance. At $\eta L = 1$ for example, the approximate acceptance is 31\% larger than the exact value with uniform aperture, and 2\% larger for an aperture at the defocussing lenses only.

We turn now to some specific examples. Consider first a beam of
ground state CaF molecules travelling at 350\,m/s into a gapless
guide having $E_{0} = 100$\,kV/cm and $a_{3}=0.2$. In this case,
$\eta = 0.053$, and with $r_{0} = 1$\,mm the acceptance reaches
its maximum value of 39\,mm\,mrad when the lenses are 24\,mm in
length. Thus, the acceptance of the guide, which can transport the
molecules over an arbitrarily long distance, is the same of that
of a 1\,mm$^2$ square cross-section pipe of length 26\,mm. As a
second example, consider Li atoms launched with a speed of 10\,m/s
from a magneto-optical trap (MOT) into the same guide. Here, $\eta
= 0.31$, the optimal lens length is 4\,mm and the transverse
acceptance is 230\,mm.mrad, equivalent to 2.3\,mm.m/s in
position-velocity space. This acceptance is larger than the
phase-space area occupied by the atoms in a typical Li MOT,
indicating that all the atoms can be successfully transported.

\subsection{Fill factor}

The previous section showed how to calculate the transverse
phase-space acceptance, but often the source is unable to fill
this acceptance area completely. The final output of particles
depends on the overlap between the acceptance and the phase-space
area occupied by the particles at the guide's entrance, which we
call the fill factor. As an illustration, consider a beam formed
in a supersonic expansion, passing through a small aperture (a
skimmer) and then into the guide. For the purposes of calculation
it is convenient to proceed in the opposite direction. We
calculate the phase-space acceptance at the entrance of the guide,
project this back through the skimmer and onto the source to find
out what fraction is filled by the source. In practice, the source
is usually large enough and divergent enough to completely fill
the part of the acceptance that fits through the skimmer, and we assume this to be the case in the calculation that follows.

\begin{figure}
\centering
\includegraphics{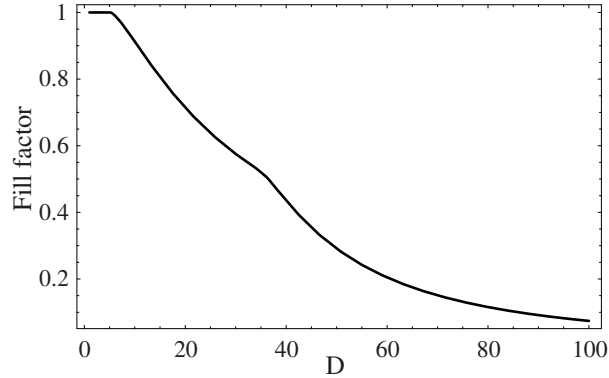}
\caption{ The fill factor as a function of the distance $D$ (in
units of $r_{0}$) between the entrance of the guide and an
upstream square aperture of side $r_{0}$. The result shown is for
the particular case of a gapless guide with $\eta=0.07$ and
$L=20$. \label{Fig:FillFactor}}\end{figure}

Let us write the phase-space ellipse accepted by the guide as

\begin{equation}
A_{1}X^{2} + 2B_{1}X X' + C_{1}X'^{2} = 1. \label{Eq:ellipse}
\end{equation}

\noindent Here, $A_{1}=\beta_{max}\gamma(0)$,
$B_{1}=\beta_{max}\alpha(0)$ and $C_{1}=\beta_{max}\beta(0)$,
where $Z=0$ is defined to be at the entrance of the guide. We
project this ellipse through a distance $-D$ and so obtain a new
ellipse with coefficients $A_{2}, B_{2}, C_{2}$. The relationship
between the coefficients is

\begin{equation}
\left( \begin{array}{c} A_{2} \\ B_{2} \\ C_{2}
\end{array} \right)= \left( \begin{array}{ccc} 1 & 0 & 0 \\
D & 1 & 0 \\
D^{2} & 2D & 1
\end{array} \right)\left( \begin{array}{c} A_{1} \\ B_{1} \\ C_{1}
\end{array} \right).
\label{Eq:ellipseDrifter}
\end{equation}

\noindent This is easily derived using the drift matrix $O(Z)$
(see (\ref{Eq:FocusMatrix})).

At this point, we can calculate the area of that part of the
phase-space ellipse that fits through the skimmer aperture. The
aperture provides a constraint in the $X$ direction, but not in
the $X'$ direction. Integrating the ellipse between $X=\pm R$, we
obtain the area \footnote{This procedure is only accurate for a
rectangular aperture, since we have assumed throughout that we can
treat the two transverse directions independently, multiplying the
results together at the end.},

\begin{equation}
\int^{R}_{-R}\frac{2}{C_{2}}\sqrt{(B_{2}^{2}-A_{2}
C_{2})X^{2}+C_{2}}\,dX.
\end{equation}

\noindent Repeating this for the other transverse direction, and
multiplying together the two results yields the 2D transverse
phase-space acceptance of the entire guide and aperture setup.
This result, divided by the 2D acceptance of the guide alone,
$(\pi/\beta_{max})^{2}$, gives the fill factor. Figure
\ref{Fig:FillFactor} shows an example of how the fill factor
decreases as $D$ increases, stressing the importance of minimizing
the distance between the guide and any aperture placed
upstream.

\section{An effective potential approach to the nonlinear dynamics}

In order to solve the nonlinear problem, whose solution is $X(Z)$,
let us define new coordinates, $w$ and $s$, in terms of the
function $\beta(z)$ which appears in the solution to the linear
problem, equation (\ref{Eq:trajEq}):

\begin{eqnarray}
w &= \frac{X(Z)}{\sqrt{\beta(Z)}},\nonumber \\
s &= \frac{1}{\mu}\int^{Z}_{0}\frac{1}{\beta(\tilde Z)}\,d\tilde Z
= \psi/\mu, \label{Eq:FloquetTransforms}
\end{eqnarray}

\noindent The phase advance $s$ increases by 1 each time the
particle advances through one unit cell. The quantity $w$ represents the macromotion of the particle. Applying the transformation to (\ref{Eq:EqMotionLinear}) we obtain

\begin{equation}
X'' + K X = \frac{1}{\mu^{2}\beta^{3/2}}\left[\frac{d^{2} w}{d s^{2}} + (1/2\,
\beta\,\beta'' - \alpha^{2} + K \beta^{2})\mu^{2}w\right] = 0,
\label{Eq:EqOfMotionTransform}
\end{equation}

\noindent where the primes still denote differentiation with respect to $Z$. Note that we
have made no assumptions about the form of $X(Z)$ in this
transformation. With the help of (\ref{Eq:diffBeta}) and
(\ref{Eq:alpha}) we see that $1/2\, \beta\,\beta'' - \alpha^{2} +
K \beta^{2} = 1$, and so our equation of motion in the transformed
coordinates reduces to that of a harmonic oscillator of angular
frequency $\mu$,

\begin{equation}
\ddot w + \mu^{2} w = 0, \label{Eq:ho}
\end{equation}

\noindent where the dots denote differentiation with respect to $s$. In the transformed coordinates the micromotion has disappeared and the macromotion appears as motion in a harmonic potential whose curvature is proportional to $\mu^{2}$.

Re-introducing the nonlinear terms, we write (\ref{Eq:EqnOfMotion}) in the form

\begin{eqnarray}
X'' + \eta^{2}(G X - G^{2}\zeta X^{3}-G^{2} \chi Y^{2}X) = 0, \nonumber \\
Y'' + \eta^{2}(-G Y - G^{2}\zeta Y^{3} - G^{2} \chi X^{2}Y) = 0,
\label{Eq:EqnOfMotionComplete}
\end{eqnarray}

\noindent where we use the function $G(Z)$ introduced earlier so
as to handle an arbitrary sequence of converging lenses, diverging
lenses and drift spaces. Applying the same coordinate
transformation as above, this becomes

\numparts
\begin{equation}
\ddot w_{1} + \mu^{2}(w_{1} - G^{2}\zeta \eta^{2}\beta_{1}^{3}
w_{1}^{3} - G^{2}\chi \eta^{2}\beta_{1}^{2} \beta_{2}
w_{2}^{2}w_{1}) = 0,\label{Eq:w1}
\end{equation}
\begin{equation}
\ddot w_{2} + \mu^{2}(w_{2} - G^{2}\zeta \eta^{2}\beta_{2}^{3}
w_{2}^{3} - G^{2}\chi \eta^{2}\beta_{2}^{2} \beta_{1}
w_{1}^{2}w_{2}) = 0. \label{Eq:w2}
\end{equation}
\endnumparts

\noindent In (\ref{Eq:w1}), $w_{1}$, $\beta_{1}$ and $s_{1}$
are identical to the $w$, $\beta$ and $s$ introduced earlier, and
the dot indicates differentiation with respect to $s_{1}$. The
quantities $w_{2}$, $\beta_{2}$ and $s_{2}$ in (\ref{Eq:w2})
are the corresponding quantities for the $y$ motion, and
differentiation is with respect to $s_{2}$. Naturally, $\beta_{1}$
and $\beta_{2}$ are related: $\beta_{2}(Z+L_{\rm cell}/2) =
\beta_{1}(Z)$.

So far, our analysis has been exact. In order to make progress,
let us assume that the guide is operated well within the range
of stability. In this regime the amplitude of the micromotion is
small compared to the macromotion amplitude and the envelope
function, $\beta(Z)$, is approximately constant. We set $\beta_{1}
= \beta_{2} = \beta_{0}$. It follows from (\ref{Eq:mu}) and
(\ref{Eq:FloquetTransforms}) that $\beta_{0} = L_{\rm cell}/\mu$
and $s_{1}=s_{2} = Z/L_{\rm cell}$. In addition, the wavelength of the macromotion is large compared to the length of a unit cell in this regime, and so $G^{2}$
oscillates rapidly relative to the macromotion wavelength. We will replace it by its mean value $\langle G^{2} \rangle$. In this way,
the same formalism will apply to guides with or without gaps. For many practical guides these approximations will turn out to be useful and not very restrictive.

Making these replacements in (\ref{Eq:w1}) and
(\ref{Eq:w2}), and then transforming back to the more familiar
variables, $X$ and $Y$, we obtain

\begin{equation}
X'' + \mu^{2} X - \epsilon b_{1} X^{3} - \epsilon b_{2} Y^{2}X =
0, \label{Eq:ReducedEqsOfMotion}
\end{equation}

\noindent where differentiation is with respect to $s$,
interchange of $X$ and $Y$ yields the second of the two equations
of motion, and

\begin{eqnarray}
\epsilon b_{1} = \zeta (\eta L_{\rm cell})^{2} \langle G^{2} \rangle,\\
\epsilon b_{2} = \chi (\eta L_{\rm cell})^{2} \langle G^{2}
\rangle.
\end{eqnarray}

\noindent Patience will reward the reader's curiosity about the
seemingly superfluous little epsilon.

In the small micromotion approximation, we can think of the
particles moving in an effective potential, $V_{\rm eff}$. From (\ref{Eq:ReducedEqsOfMotion}), and the equation obtained by
interchange of $X$ and $Y$, we find this effective potential to be
given by

\begin{equation}
V_{\rm eff}/M = \case{1}{2} \mu^{2}(X^{2} + Y^{2}) -
\case{1}{4}\epsilon b_{1}(X^{4}+Y^{4}) - \case{1}{2}\epsilon
b_{2} X^{2}Y^{2}.\label{Eq:Veff}
\end{equation}

\noindent We also note that the transverse phase space acceptance obtained by applying the small micromotion approximation in the harmonic case is simply $\pi \mu$. Converting from our present coordinates, $(X,\frac{dX}{ds})$, back to $(x,\frac{dx}{dz})$, this acceptance becomes $\pi \mu\,\eta\,r_{0}/(\eta L_{\rm cell})$. This result was already discussed below (\ref{Eq:exampleAcceptance}) in the context of figure \ref{Fig:LinAcc}.

\section{The cubic term}

In this section we derive analytical expressions for the
trajectories of the particles and for the phase space acceptance
of the guide, with the cubic term included. We set $b_{2} = 0$ in (\ref{Eq:ReducedEqsOfMotion}), multiply by $X'$, and then integrate with respect
to $s$, obtaining

\begin{equation}
\case{1}{2} X'^{2} + \case{1}{2}\mu^{2}X^{2} - \case{1}{4}
\epsilon b_{1} X^{4} = h, \label{Eq:cubicEnergy}
\end{equation}

\noindent where $h$ is a constant of the motion (the total
energy), and we identify the first term as the kinetic energy and
the second and third terms as the potential energy. Rearranging
for $X'$ and integrating gives

\begin{equation}
s = \frac{1}{\sqrt{2h}} \int\frac{dX}{\sqrt{1-p X^{2} +q X^{4}}},
\label{Eq:sFuncX}
\end{equation}

\noindent where $p=\mu^{2}/2h$ and $q=\epsilon b_{1}/4h$. To
complete the solution, we factorize the expression under the
square root to the form $(1-a_{+} X^{2})(1- a_{-} X^{2})$ where
$a_{\pm} = p/2 \pm \sqrt{p^{2}/4 - q}$. Then, the integral takes
the standard form for an elliptic integral of the first kind, $F$,
and the solution is\footnote{Conventions for elliptic integrals
are not standardized. We use the convention where $F(\phi,m) =
\int_{0}^{\phi}(1-m\sin^{2}(\theta))^{-1/2}\,d\theta$. The inverse
of this elliptic integral is the Jacobi amplitude $am(u,m)$ such
that if $u=F(\phi,m)$, then $\phi = am(u,m)$. The Jacobi elliptic
function is defined by $sn(u,m) = \sin(am(u,m))$.}

\begin{equation}
s(X) = \frac{1}{\sqrt{2a_{+} h}}\,F(\sin^{-1}(\sqrt{a_{+}} X),
a_{-}/a_{+}).
\end{equation}

\noindent Finally, this expression can be inverted to give $X(s)$ in terms
of a Jacobi elliptic function:

\begin{equation}
X(s) = \frac{1}{\sqrt{a_{+}}}\,{\rm sn}(\sqrt{2a_{+} h }\,s,
a_{-}/a_{+}).\label{Eq:cubicTraj}
\end{equation}

\begin{figure}
\centering
\includegraphics{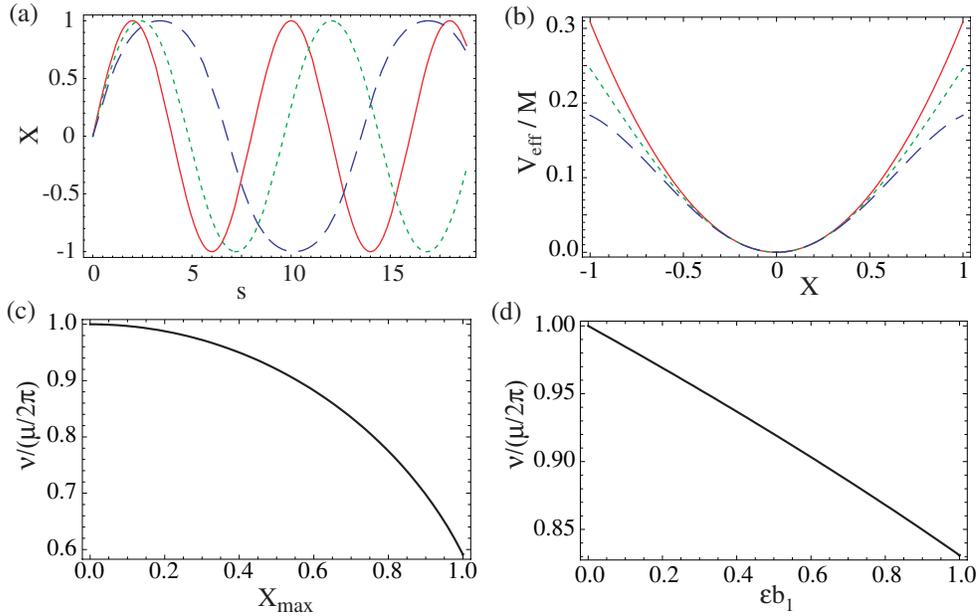}
\caption{(a) The macromotion for the case $\mu=\pi/4$ and
$X_{{\rm max}}=1$ and for three different values of $\epsilon b_{1}$.
Solid line: $\epsilon b_{1} = 0$. Dotted line: $\epsilon b_{1} =
0.25$. Dashed line: $\epsilon b_{1} = 0.5$. (b) Plots of the effective potential along the $X$ axis, when $b_{2}=0$, $\mu = \pi/4$, and $\epsilon b_{1}$ has the same three values as in (a). The equation for the effective potential is (\ref{Eq:Veff}). (c) Macromotion spatial frequency as a function of $X_{{\rm max}}$, when $\epsilon b_{1} = 0.5$. (d)  Macromotion spatial frequency as a function of $\epsilon b_{1}$, when $X_{{\rm max}}=0.5$.
In (c) and (d) we have used the exact expression for $\nu$ given by (\ref{Eq:Lambda}), normalized to $\nu_{0}=\mu/(2\pi)$.\label{Fig:Macromotion}}\end{figure}

\noindent This equation provides an analytical form for the
trajectory of a particle through the guide, in the presence of the
cubic nonlinearity, but averaged over the micromotion. Figure
\ref{Fig:Macromotion}(a) shows some examples of how the
trajectories depend on the size of the cubic term, calculated
using (\ref{Eq:cubicTraj}). The figure shows that as
$\epsilon b_{1}$ increases, the wavelength increases and the
trajectories evolve from a sinusoidal shape towards a more square
shape. These changes reflect the way the effective potential
changes with $\epsilon b_{1}$. Part (b) of the figure shows the effective potential along the $X$-axis for the same parameters used to obtain the trajectories plotted in (a). As the value of $\epsilon b_{1}$ increases, the steepness of the effective potential well
decreases, and so the wavelength of the macromotion increases. In
the extreme case where $\epsilon b_{1} = \mu^{2}$, the particle
will stop at the maximum of the potential and remain there
forever.

From (\ref{Eq:sFuncX}) and the discussion that follows it, we
find the wavelength of the macromotion to be

\begin{equation}
\Lambda = \frac{4}{\sqrt{2h}}
\int_{0}^{X_{{\rm max}}}\frac{dX}{\sqrt{1-a_{+} X^{2}} \sqrt{1 - a_{-}
X^{2}}},
\end{equation}

\noindent $X_{{\rm max}}$ being the turning point of the motion where
$X'=0$, found to be $X_{{\rm max}}=1/\sqrt{a_{+}}$. The wavelength
evaluates to

\begin{equation}
\Lambda = \frac{4}{\sqrt{2h a_{+}}}
K(a_{-}/a_{+}),\label{Eq:Lambda}
\end{equation}

\noindent where $K$ is the complete elliptic integral of the first
kind. It is illuminating to write down the series expansion for
$\Lambda$, or its inverse, the spatial frequency $\nu$, in powers
of the small quantity $\epsilon b_{1}$:

\begin{equation}
\nu = \frac{1}{\Lambda} = \frac{\mu}{2\pi} \left(1-\frac{3\epsilon
b_{1} X_{{\rm max}}^{2}}{8\mu^{2}} - \frac{21 \epsilon^{2} b_{1}^{2}
X_{{\rm max}}^{4}}{256 \mu^{4}} + \ldots
\right).\label{Eq:FrequencyExpansionCubic}
\end{equation}

\noindent For positive $\epsilon b_{1}$, the spatial frequency
decreases with increasing $\epsilon b_{1}$ and with increasing
oscillation amplitude. To lowest order, the change depends on the
square of the oscillation amplitude. Particles that have a large
amplitude will oscillate more slowly through the guide than those
that remain close the guide's axis. Figure
\ref{Fig:Macromotion}(c) and (d) illustrate the exact spatial frequency
versus $X_{{\rm max}}$ and $\epsilon b_{1}$. These plots shows the
quadratic dependence on the oscillation amplitude and the linear
dependence on $\epsilon b_{1}$.

We now calculate the phase-space acceptance of the guide in the
presence of the cubic term. The two transverse directions are uncoupled at this point in our discussion, so it is convenient to calculate the acceptance in 1D, squaring the result to get the total 2D acceptance. When $\epsilon b_{1}$ is positive, the
effective potential in which the particles move has a turning
point at $X=X_{c}$ where $\mu^{2}X_{c} - \epsilon b_{1} X_{c}^{3}
= 0$, i.e.

\begin{equation}
X_{c} = \sqrt{\frac{\mu^2}{\epsilon b_{1}}}. \label{Eq:wc}
\end{equation}

\noindent Particles with enough energy to reach this turning point
cannot be confined in the guide. A particle that has just enough
energy to reach the top of the effective potential well has an
energy $h_c = 1/2 \mu^{2} X_{c}^{2} - 1/4 \epsilon b_{1} X_{c}^4 =
\mu^{4}/(4\epsilon b_{1})$. To determine the phase-space
acceptance we need to distinguish between two cases. In the first
case, the turning point is `inside' the physical aperture and so
the acceptance is limited by the turning point. In the second
case, the turning point is `outside' the physical aperture, or
there is no turning point at all (because $\epsilon b_{1} < 0$),
and the acceptance is limited by the physical aperture. In the
first case, the equation defining the area in phase space where
particles stay inside the potential is

\begin{equation}
\case{1}{2} X'^{2} + \case{1}{2} \mu^{2} X^{2} - \case{1}{4} \epsilon b_{1} X^{4} =
\mu^{4}/(4\epsilon b_{1}).
\end{equation}

\noindent The enclosed area is the phase space acceptance and is easily
calculated by rearranging for $X'$ and integrating under the curve
from $X=0$ to $X=X_c$ where the curve crosses the $X$-axis. The
result is

\begin{equation}
A = 4\mu \int_{0}^{X_c}
\sqrt{\frac{X_{c}^{2}}{2}-X^{2}+\frac{X^{4}}{2X_{c}^{2}}}\,dX =
\frac{4\sqrt{2}\mu^{3}}{3\epsilon b_{1}},\label{Eq:AreaCase1}
\end{equation}

\noindent where we have used equation (\ref{Eq:wc}). The
phase-space acceptance scales as the cube of the phase-advance per
unit cell, a measure of the trapping strength, and is inversely proportional to $\epsilon b_{1}$, the coefficient of the nonlinearity.

In the second case, where the acceptance is limited by the
physical aperture at $X=X_{A}$, the equation defining the accepted
phase-space area is

\begin{equation}
\case{1}{2}X'^{2}+\case{1}{2}\mu^{2}X^{2}-\case{1}{4}\epsilon b_{1} X^{4} = \case{1}{2}\mu^{2}X_{A}^{2}-\case{1}{4}
\epsilon b_{1} X_{A}^{4}.
\end{equation}

\noindent By integrating the expression for $X'$ with respect to
$X$ from $X=0$ to $X=X_{A}$, we find the area enclosed by this
curve to be

\begin{equation}
A=\frac{8}{3}\mu X_{c}^{2}\sqrt{1-1/2(X_{A}/X_{c})^{2}}\{E(\kappa)
-[1-(X_{A}/X_{c})^{2}]K(\kappa)\},\label{Eq:AreaCase2}
\end{equation}

\noindent where $\kappa = X_{A}^{2}/(2X_{c}^{2}-X_{A}^{2})$, and
$K$ and $E$ are the complete elliptic integrals of the first and
second kinds. We note that, for $X_{A}=1$, this equation reduces to $A=\pi \mu$ in the limit where $\zeta \rightarrow 0$. This is the result we already found when applying the constant beta approximation in the harmonic limit. We find it useful to factor out the $\pi \mu$, writing our result as the product of the acceptance in the harmonic limit, $A_{0}$, and a factor, $F_{\zeta}$, that accounts for the cubic nonlinearity:

\numparts
\begin{equation}
A = A_{0}F_{\zeta},
\end{equation}
where
\begin{equation}
F_{\zeta} = \frac{4\sqrt{2}}{3 \pi} X_{c}^{2}\label{Eq:cubicF1}
\end{equation}
when $X_{c} \le X_{0}$, and
\begin{equation}
F_{\zeta} = \frac{8}{3\pi}X_{c}^{2}\sqrt{1-1/2(X_{A}/X_{c})^{2}}\{E(\kappa)
-[1-(X_{A}/X_{c})^{2}]K(\kappa)\}\label{Eq:cubicF2}
\end{equation}
\endnumparts

\noindent when $X_{c} > X_{0}$. By writing the result in this way, we can choose to use the exact result for $A_{0}$, derived using the procedure detailed in section \ref{Sec:LinearPSA}, rather than $\pi \mu$ which we already know from figure \ref{Fig:LinAcc} is not very accurate unless $\eta L$ is small. In this way, we effectively use the small micromotion approximation to handle only the nonlinear part of the problem. This approach, which we apply throughout, is justified by the high accuracy of the results it gives when compared to numerical simulations.

Together, (\ref{Eq:cubicF1}) and (\ref{Eq:cubicF2}) give the phase-space acceptance for all values of $\epsilon b_{1}$. The
accuracy of this analytical approach depends on the validity of
the approximation we have made, namely that $\beta$ is
approximately constant. To explore this, we can calculate the
acceptance numerically for some specific cases, and then see how
well the numerical and analytical results compare. To calculate
the acceptance exactly, we set $\chi = 0$ in (\ref{Eq:EqnOfMotionComplete}) and then solve this equation
numerically for a large number of particles, taking care to ensure
that the initial phase-space distribution is large enough to
completely fill the acceptance area. Particles whose trajectories
exceed the transverse boundaries of the guide (at $X=\pm1$), are
removed from the calculation so that we are left with the set of
particles transmitted by the guide. We calculated for a guide consisting of 200 lenses. This was long enough to show that the phase-space acceptance is independent of the guide's length, provided this is longer than the wavelength of the macromotion.

\begin{figure}
\centering
\includegraphics{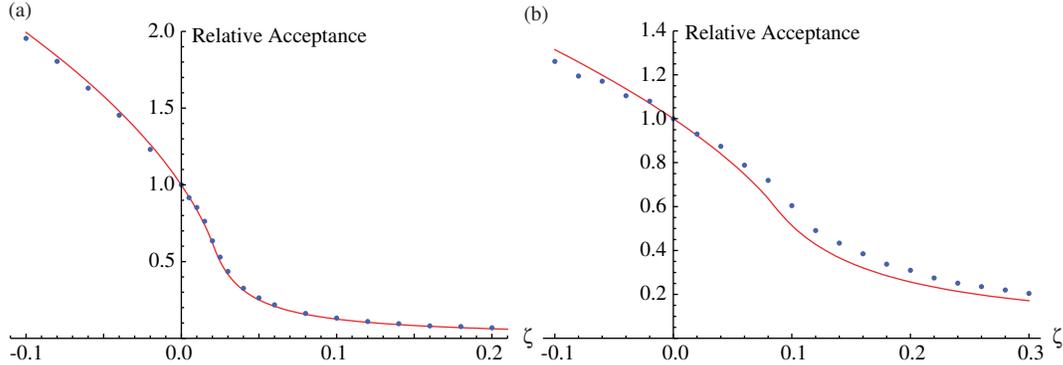}
\caption{The effect of the cubic term, characterized by $\zeta$, on the 1D acceptance of a gapless guide. Points and lines
are numerical and analytical results for (a) $\eta L = 0.5$, and
(b) $\eta L = 1$.\label{Fig:Cubic}}\end{figure}

Figure \ref{Fig:Cubic} shows how the acceptance of a gapless guide
depends on the value of $\zeta$. It compares the exact acceptance,
calculated numerically, with the analytical result, $F_{\zeta}$, given by (\ref{Eq:cubicF1}) and (\ref{Eq:cubicF2}). Part (a) shows the results obtained with
$\eta L = 0.5$ for which the phase advance per cell is $\mu =
0.14$. The numerical results have been normalized to the exact acceptance obtained when
$\zeta = 0$, which in this case is $0.426\eta r_{0}$, while the analytical result is already normalized. When $\zeta$
is positive, the acceptance decreases rapidly with increasing
$\zeta$, as the turning point of the effective potential moves
inwards. For negative values of $\zeta$, the aberration increases
the curvature of the effective potential and so increases the
acceptance. The analytical and numerical results show excellent
agreement for all values of $\zeta$ calculated. For positive
$\zeta$ the results agree to within the error bars on the
numerical values, which are approximately the size of the points.
For negative values of $\zeta$, the analytical result is
consistently 2\% larger than the numerical results. Part (b) of
the figure shows the results obtained with $\eta L = 1$ for which
$\mu=0.58$. Again, the numerical results have been normalized to the
aberration-free acceptance, which is $0.701\eta r_{0}$. Here, our
small micromotion approximation begins to break down, and the
true acceptance is less well represented by the analytical result.
Nevertheless, the analytical result differs by 20\% at most from
the true value. The analytical approximation is very useful since it can be evaluated instantly, unlike the numerical simulation. The full 2D acceptance is simply the
square of the 1D acceptance that we've calculated.

We now examine the effect of the cubic term on some example
guides. At present, all the experimental work on alternating
gradient deceleration of polar molecules has used two-rod lenses.
In these decelerators, two rods, 6\,mm in diameter, with centres separated by 8\,mm, are
aligned with their axes parallel to the molecular beamline, and
the experiments are performed in the linear Stark shift regime. In
this case $a_{5}/a_{3} = a_{3} = 1/7$ and so $\zeta = 2/7$.
Usually, the length, $S$, of the drift spaces is approximately
equal to the length of the lenses, and so we take $\eta S = \eta
L$. We use our analytical formulae to calculate how the cubic term
reduces the 2D phase space acceptance in this geometry,
normalizing the results to the values obtained when $\zeta = 0$.
When $\eta L =0.5$, the normalized acceptance, $F_{\zeta}$, is a miserable 3\%.
This grows to 18\% when $\eta L = 0.75$ and to 60\% when $\eta L =
1$. We see that the nonlinearity is hugely detrimental if the
guide is operated at low values of $\eta L$ where the phase
advance per cell, and hence the depth of the effective potential,
is small. As $\eta L$ is increased towards the edge of stability, the
nonlinearity becomes less important. This geometry is even worse
for the electric guiding of cold atoms whose Stark shift is
quadratic, because the quadratic Stark shift gives $\zeta = 3
|a_{3}|$ instead of $2|a_{3}|$.

In the four rod guide considered in \cite{Bethlem(1)06}, rods of radius $2.3r_0$ were placed with centres on the corners of a square of side $6.6r_0$. The rods on one side of the square were at positive high voltage, and those on the other side at negative high voltage, so the arrangement has two-pole symmetry. In this arrangement the cubic term is small when the Stark shift is linear, $\zeta=0.028$, and the normalized acceptance, $F_{\zeta}$, is 20\% when $\eta L = 0.5$ and $79\%$ when $\eta L = 1$. The guide is not so suitable when the Stark shift is quadratic, for then $\zeta = 0.17$. The structure used in \cite{Junglen04} to guide slow molecules from an effusive source was built from four rods of radius $r_0$ with centres on the corners of a square of side $3r_0$. The rods on two diagonally opposite corners were at high voltage, $\pm V$, with the other two grounded. Fitting the electrostatic potential obtained in this geometry to the multipole expansion, (\ref{Eq:phi}), we obtain $a_{3}=0.59$ and $a_{5}/a_{3}=-0.055$. So this geometry has $\zeta=-0.11$ for polar molecules with linear Stark shifts, meaning that the cubic term enhances the acceptance over the perfect linear case. However, the large value of $a_{3}$ means that our series expansion of the electric field, equation (\ref{Eq:E}), converges very slowly. The expansion does not provide an accurate representation of the field unless terms of higher order are included and so our conclusion about the effect of the cubic term is not so helpful in this case. We will see later that the lowest order transverse coupling term is severely detrimental for this geometry.

\section{Multiple scales analysis}

In the preceding section we found analytical solutions for the motion
in the presence of the cubic nonlinearity. We can no longer do so when $b_{2} \ne 0$ in equation (\ref{Eq:ReducedEqsOfMotion}). Nevertheless, provided the nonlinear terms
are small, we can elucidate the motion of the particles using a
perturbation technique known as multiple scales analysis. We are particularly concerned with stable motion in the guide, i.e. with those particles that remain inside the guide indefinitely. We recognize that the coupling term will lead to an exchange of energy between the $x$ and $y$ oscillations, and so the oscillation amplitudes will be modulated. If $\epsilon b_{2}$ is small enough, this modulation will occur on a much longer scale than the wavelength of the macromotion. To handle this, we look for a solution in the form of a perturbation expansion using two separate length scales. We follow references \cite{ManevitchBook, NayfehBook, BenderOrszagBook}, and refer the reader to these books for more details on the multiple scales method.

The independent variable in (\ref{Eq:ReducedEqsOfMotion}) is
$s = Z/L_{\rm cell}$. We introduce a second length scale, $\sigma
= \epsilon s$, treat $X$ and $Y$ as functions of both $s$ and
$\sigma$, and then expand $X$ and $Y$ as a power series in
$\epsilon$: $X(s, \sigma) = X_{0}(s,\sigma) + \epsilon
X_{1}(s,\sigma) + \ldots$. In order that the initial conditions on
$X$ and $Y$ be satisfied uniquely in this expansion, we choose to
set $X_{n}(0)=Y_{n}(0)=X'_{n}(0)=Y'_{n}(0)=0$ for all $n \ne 0$.
The derivatives are

\begin{eqnarray}
\frac{d X}{d s} &= \frac{\partial X_{0}}{\partial s} +
\epsilon\left(\frac{\partial X_{0}}{\partial \sigma} +
\frac{\partial X_{1}}{\partial s}\right) +
O(\epsilon^{2}),\nonumber\\
\frac{d^2 X}{d s^2} &= \frac{\partial^{2} X_{0}}{\partial s^{2}} +
\epsilon\left(2\frac{\partial^{2} X_{0}}{\partial s\partial
\sigma} + \frac{\partial^{2} X_{1}}{\partial s^{2}}\right) +
O(\epsilon^{2}).\nonumber
\end{eqnarray}

\noindent Substituting into (\ref{Eq:ReducedEqsOfMotion}) we
obtain a differential equation whose solution should not depend on
the arbitrary quantity $\epsilon$. We therefore have a separate
equation for each power of $\epsilon$, these being

\begin{eqnarray}
\frac{\partial^{2} X_{0}}{\partial s^{2}} + \mu^{2} X_{0} &=
0, \label{Eq:X0Diff}\\
\frac{\partial^{2}X_{1}}{\partial s^{2}} + \mu^{2} X_{1} &=
-2\frac{\partial^{2}X_{0}}{\partial s \partial \sigma} + b_{1}
X_{0}^{3} + b_{2}Y_{0}^{2}X_{0},\label{Eq:X1Diff}
\end{eqnarray}

\noindent and similarly for $Y_{0},Y_{1}$, by simple interchange
of the symbols $X$ and $Y$.

The solutions for $X_{0}$ and $Y_{0}$ are

\begin{eqnarray}
X_{0} = \frac{r_{1}}{2} e^{i(\mu s + \theta_{1})} + c.c,\nonumber\\
Y_{0} = \frac{r_{2}}{2} e^{i(\mu s + \theta_{2})} + c.c,
\label{Eq:lowestOrderSolution}
\end{eqnarray}

\noindent where $c.c$ stands for the complex conjugate and
$r_1,r_2,\theta_1,\theta_2$ are real functions of $\sigma$, but
independent of $s$. These solutions are now substituted into
(\ref{Eq:X1Diff}), giving, after a little algebra,

\begin{eqnarray}
&\frac{\partial^{2} X_1}{\partial s^{2}} + \mu^{2}X_{1} = c_1 e^{i\mu s} +
c_3 e^{3i\mu s} + c.c, \nonumber\\
&c_{1}= \left(-i \mu \frac{d r_1}{d \sigma }+\mu r_1\frac{d \theta
_{1}}{d \sigma} + \frac{3b_{1}}{8} r_{1}^{3} +
\frac{b_{2}}{4}r_{1}r_{2}^{2}\right)e^{i\theta_1} + \frac{b_2}{8}r_{1}r_{2}^{2}e^{i(2\theta_{2} - \theta_{1})},\nonumber\\
&c_3 = \frac{b_1}{8}r_{1}^{3}e^{3i\theta_{1}} +
\frac{b_2}{8}r_{1}r_{2}^{2}e^{i(2\theta_{2}+\theta_{1})}.
\label{Eq:X1Diff2}
\end{eqnarray}

Consider for a moment the solution to the above differential
equation with the right hand side replaced by zero (the
corresponding homogeneous equation). Its solution has terms in
$e^{i \mu s}$ and $e^{-i \mu s}$, terms which also appear in the
right hand side of the inhomogeneous equation, (\ref{Eq:X1Diff2}). Terms in an inhomogeneous equation which
are themselves solutions to the associated homogeneous equation
lead to secular terms in the solution. These secular terms grow
with $s$ more rapidly than the corresponding solution of the
homogeneous equation by at least a factor of $s$ \cite{BenderOrszagBook}, i.e. as $s
\cos(\mu s)$ in this case. If
our solution is to represent stable motion, corresponding to
particles trapped in the guide indefinitely, we must have to
eliminate the secular terms by setting to zero the coefficient,
$c_1$, of $e^{i\mu s}$ in (\ref{Eq:X1Diff2}). We are free to do this, since the $r$'s and $\theta$'s are otherwise undetermined functions of $\sigma$. Setting $c_{1}=0$ and then separating out the real and imaginary parts, we obtain the following
differential equations describing the slow variation of $r_{1},
r_{2}, \theta_{1}, \theta_{2}$:

\numparts
\begin{equation}
r_{1}\frac{dr_{1}}{d\sigma }-\frac{b_2}{8\mu
}r_{1}^{2}r_{2}^{2}\sin (2(\theta_{2}-\theta_{1}))=0, \label{Eq:slowDiffs1}
\end{equation}
\begin{equation}
r_{2}\frac{dr_{2}}{d\sigma }+\frac{b_2}{8\mu
}r_{2}^{2}r_{1}^{2}\sin (2(\theta_{2}-\theta_{1}))=0, \label{Eq:slowDiffs2}
\end{equation}
\begin{equation}
\frac{d\theta _1}{d\sigma }+\frac{3b_1}{8\mu
}r_1{}^2+\frac{b_2}{8\mu }r_2{}^2[2+\cos (2(\theta_{2}-\theta_{1}))]=0, \label{Eq:slowDiffs3}
\end{equation}
\begin{equation}
\frac{d\theta _2}{d\sigma }+\frac{3b_1}{8\mu
}r_2{}^2+\frac{b_2}{8\mu }r_1{}^2[2+\cos (2(\theta_{2}-\theta_{1}))]=0. \label{Eq:slowDiffs4}
\end{equation}
\endnumparts

By solving this set of differential equations we obtain the slow
evolution of the amplitudes and phases, which we can then
substitute into equations (\ref{Eq:lowestOrderSolution}) to get the
full solution to lowest order in $\epsilon$. As we will see later,
this procedure is remarkably accurate. A great deal of insight
into the motion of the particles can be acquired from the above
set of equations. By adding together (\ref{Eq:slowDiffs1}) and
(\ref{Eq:slowDiffs2}) we see that $r_{1}^{2} + r_{2}^{2} = E_{0}$,
a constant, proportional to the total energy at this order of
approximation. Subtracting (\ref{Eq:slowDiffs1}) from
(\ref{Eq:slowDiffs2}) and (\ref{Eq:slowDiffs3}) from
(\ref{Eq:slowDiffs4}) yields

\begin{eqnarray}
\frac{d\xi }{d\sigma }-\frac{b_2}{4\mu }E_0\xi (1-\xi )\sin
(2\gamma )=0, \label{Eq:Diffxi}\\
\frac{d\gamma }{d\sigma }+E_0(1-2\xi )\left(\frac{3b_1}{8\mu
}-\frac{b_2}{8\mu }(2+\cos (2\gamma ))\right)=0,
\label{Eq:Diffgamma}
\end{eqnarray}

\noindent where $\xi$ is the fractional energy associated with
oscillations in the $x$ direction, $\xi=r_{1}^{2}/E_{0}$, and
$\gamma = \theta_{2} - \theta_{1}$ is the phase difference between
the two transverse oscillations. The solutions of these equations
tell us how the phase difference and the partitioning of the
energy evolve with the axial coordinate. Even more instructive is
to eliminate the independent variable by dividing
(\ref{Eq:Diffgamma}) by (\ref{Eq:Diffxi}), and then integrating to
give

\begin{equation}
\{3b_1-b_2[2+\cos (2\gamma )]\}\xi (\xi -1)=C,
\label{Eq:xigammacurves}
\end{equation}

\noindent where $C$ is a constant. This equation is the central
result of this analysis. It tells us that there is an invariant
quantity involving only $\xi$ and $\gamma$. For a given particle,
the value of $C$ is fixed by the the initial conditions. The
particle is then confined to travel on the contour in
$(\xi,\gamma)$ space defined by the above equation. This tells us how the amplitudes of the two transverse oscillations will change,
allowing us to determine whether a particle will remain in the
guide or crash into its boundaries.

\section{The nonlinear coupling}

\subsection{Small coupling coefficient}\label{Sec:smallCoupling}

\begin{figure}
\centering
\includegraphics{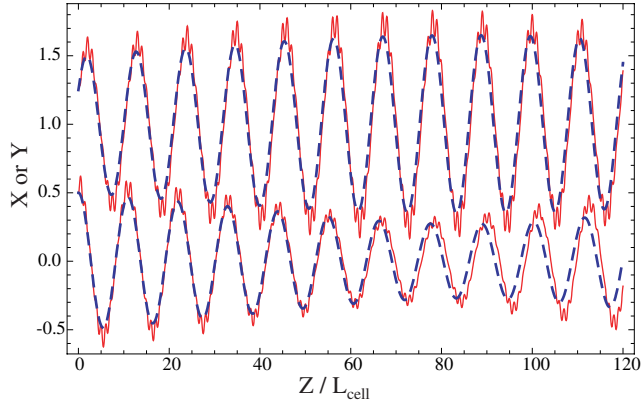}
\caption{Trajectories through an alternating gradient guide with
nonlinear transverse coupling. The guide has $\eta L = 1$, $\eta S
= 0$, $\chi = 0.03$. The initial conditions of the particle were
$X(0)=0.5,\,X'(0)=0,\,Y(0)=0.25,\,Y'(0)=0.25$. Solid line: Exact
trajectory calculated numerically. Dashed line: Approximate
trajectory given by (\ref{Eq:lowestOrderSolution}) and the
solutions to (\ref{Eq:slowDiffs1}) - (\ref{Eq:slowDiffs4}). The $Y$-trajectory has
been offset (by 1 unit) for
clarity.\label{Fig:CoupledTrajectories}}
\end{figure}

In this section, we use the results obtained above to understand
the effect of the nonlinear coupling on the motion of particles in
the guide. To isolate the effect of the coupling, we set $b_{1} =
0$. To demonstrate that the approximation developed above is a
useful one, we begin by looking at an example trajectory. The
solid line in Figure \ref{Fig:CoupledTrajectories} shows the exact
trajectory of a particle travelling through 120 unit cells of a
gapless guide that has $\eta L = 1$, $\zeta = 0$ and $\chi =
0.03$. The trajectory shows the usual superposition of micromotion
and macromotion. The transverse coupling is responsible for the slow
modulation of amplitude and relative phase evident in the
figure. The dashed line in the figure shows the result of the multiple scales
approximation. The parameters given above correspond to
$\mu = 0.585$ and $\epsilon b_{2} = (\eta L_{\rm cell})^{2} \chi =
0.12$, and the dashed curve is obtained from
(\ref{Eq:lowestOrderSolution}) using the solutions to (\ref{Eq:slowDiffs1}) - (\ref{Eq:slowDiffs4}). We see that the approximate solution
does a good job of describing the true motion of the particle,
including the modulation of the amplitude and phase. In fact, in
order that the approximate and exact trajectories could be
distinguished in the figure, we had to choose a case where our
approximations begin to break down - the micromotion amplitude is
not so small relative to the macromotion, and $\epsilon b_{2}$ is
not very small compared to $\mu^{2}$. In situations where the
approximations hold more accurately, the two trajectories are
virtually indistinguishable.

\begin{figure}
\centering
\includegraphics{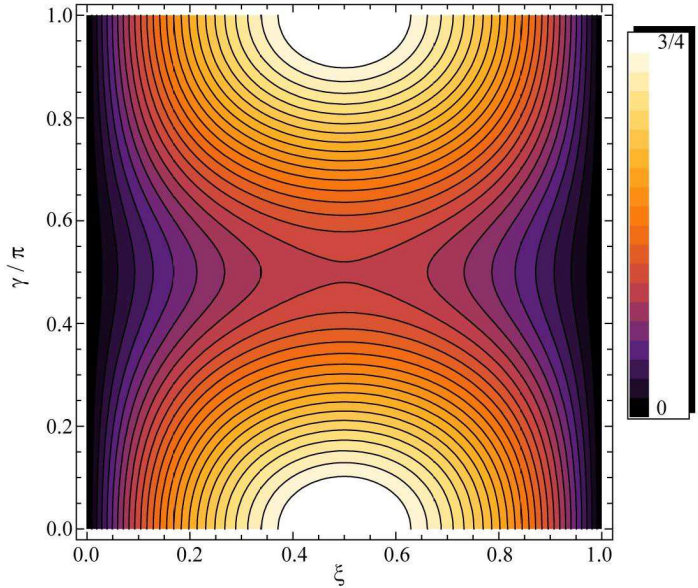}
\caption{Contour plot of (\ref{Eq:xigammacurvesZerob1})
illustrating the slow oscillations in amplitude and phase. The colour map shows the values of the contours.
\label{Fig:Contours}}
\end{figure}

Now that we have illustrated the usefulness of this approximate
approach, we continue with our analysis, returning to the key
result given in (\ref{Eq:xigammacurves}). When $b_{1}=0$,
this equation reduces to

\begin{equation}
(2+\cos (2\gamma ))\xi (1-\xi )=C/b_2,
\label{Eq:xigammacurvesZerob1}
\end{equation}

\noindent representing a set of contours of a form that is independent of $b_2$ (except as an over-all scaling factor). Figure \ref{Fig:Contours} is a plot of these contours along with a colour map indicating the values of $C/b_{2}$. Notable features in this plot are the two maxima at $(\xi,\gamma)=(\case{1}{2},0)$ and $(\case{1}{2},\pi)$, where $C/b_{2} = 3/4$, and the saddle point at $(\xi,\gamma)=(\case{1}{2},\case{\pi}{2})$ where $C/b_{2}=1/4$. The maxima and saddle point are fixed points representing coupled motion where the amplitudes and phases do not change with time - {\it stationary coupled modes}.

The two maxima correspond to motion along the diagonals of the guide. The oscillations in $x$ and $y$ have equal amplitudes and are either exactly in phase ($\gamma=0$), or exactly in anti-phase ($\gamma = \pi$). Even though the coupling is nonlinear, we have found the {\it normal modes} familiar from the theory of linearly coupled harmonic oscillators. Indeed, applying the multiple scales method to a pair of linearly coupled harmonic oscillators results in the relationship $(1+\cos (2\gamma ))\xi (1-\xi ) = {\rm constant}$, which is remarkably similar to (\ref{Eq:xigammacurvesZerob1}). The frequencies of the normal modes can be found by differentiating their phase, $\phi(s)$, with respect to $s$. The phase of the $X$ oscillation at position $s$ is, according to (\ref{Eq:lowestOrderSolution}), $\phi(s) = \mu s +
\theta_1(\sigma)$, and so the oscillation frequency is $\omega = d\phi/ds
= \mu + \epsilon\,d\theta_{1}/d\sigma$. With the help of (\ref{Eq:slowDiffs3}) the normal mode frequencies are
found to be

\begin{equation}
\omega_{\rm NM} = \mu\left(1 - \frac{3r_{1}^{2}\epsilon b_{2}}{8
\mu^2}\right). \label{Eq:FrequencyCoupled}
\end{equation}

\noindent We see that the frequency is lowered when $\chi$ (and
hence $\epsilon b_2$) is positive because the coupling term
reduces the trap depth along the diagonals of the guide. The
opposite is true when $\chi$ is negative. Note that the normal modes with $\gamma =0$ and $\gamma = \pi$ have equal frequencies, unlike the linear coupling case where the coupling removes the frequency degeneracy.  There is another important difference too: the
frequency of the normal mode differs from $\mu$ by an amount that
depends on the square of the oscillation amplitude (the total
energy). This is similar to the frequency shift caused by the
cubic term discussed earlier, and quite different from a linear
coupling whose normal mode frequencies are independent of
amplitude.

The two maxima in figure \ref{Fig:Contours} are encircled by contours with $1/4<C/b_{2}<3/4$. In this range, the two transverse oscillations exchange energy completely. Looking at the contours near $\gamma = 0$, the phase difference is zero when the amplitude of one mode has its maximum value, zero again when it has its minimum value, and reaches a maximum when the amplitudes of the two modes are equal. The phase difference never exceeds $\pi/2$. Again, this type of coupled motion is similar to that of linearly coupled oscillators. For a given value of $C/b_{2}$ in this range, the maximum value of $\xi$ is found at $\gamma = 0$ and is

\begin{equation}
\xi _{m1}=\frac{1}{2}\left(1+\sqrt{1-4C/(3b_2)}\right).\label{Eq:xiMax1}
\end{equation}

The saddle point in figure \ref{Fig:Contours} is a coupled stationary mode corresponding to circular motion in the $x y$ plane - we call it a {\it circular mode}. Its stationary nature is unstable in the sense that an infinitesimally small deviation from this point results in oscillations with time-varying amplitudes and phases. We will show later that the circular modes can become stable when $b_{1} \ne 0$.

When $C/b_{2}<1/4$ the contours are below the saddle point and so no longer encircle the maxima. In this range, although the amplitude of one mode does grow at the expense of the other, the amplitudes do not exchange completely. The contours are open and the phase difference increases continually. Of particular interest is the fact that motion with $\xi=0$ or $\xi=1$ is of this type, and does not yield any exchange of amplitude at all. The motions uncouple when all the energy is in one direction, in stark contrast to the classic motion of linearly coupled oscillators which exchange energy fully even when one oscillator begins at rest. For a given value of $C/b_{2}<1/4$, the large $\xi$ contour has maxima at $\gamma = 0$ and $\pi$, whose value, $\xi_{m1}$, is given by (\ref{Eq:xiMax1}). The small $\xi$ contour has a maximum at $\gamma = \pi/2$ with value

\begin{equation}
\xi_{m2}=\frac{1}{2}\left(1-\sqrt{1-4C/b_2}\right).\label{Eq:xiMax2}
\end{equation}

Due to the exchange of energy between the transverse modes, there
will be some particles whose maximal amplitudes of oscillation
take them outside the boundaries of the guide, even though their
initial amplitudes are within these boundaries. These particles
are lost from the guide in the presence of coupling, and would not
have been in its absence. We note again that the value of
$\epsilon b_{2}$ dictates the rate at which the amplitudes
exchange, but not the degree of exchange. If the guide is
infinitely long, and $\epsilon b_{2}$ is small enough that our
approximations hold, the loss of acceptance is independent of the
size of $\epsilon b_2$. A larger value only results in the
particles being lost more quickly, but those same particles will
eventually be lost even when the coupling term is infinitesimally
small. Similarly, and somewhat counter-intuitively, the same
particles are lost whether the nonlinear term increases or
decreases the depth of the guide (along the diagonals).

At the present level of approximation, it is straightforward to determine whether a particle with given initial conditions will be transmitted through the guide. We first relate the invariant quantity, $C/b_2$, to the initial conditions. To lowest order in $\epsilon$, the initial values of $a_1,\theta_1$ are related to the initial conditions via

\begin{eqnarray}
r_{1}(0)^{2} = X(0)^{2} + X'(0)^{2}/\mu^{2} \\
\theta_{1}(0) = \tan^{-1}\left(-\frac{X'(0)}{\mu X(0)}\right),
\end{eqnarray}

\noindent and similarly for $a_2,\theta_2$. With these, we can form $\xi(0)$ and $\gamma(0)$ and hence $C/b_{2}$ from (\ref{Eq:xigammacurvesZerob1}). An equivalent approach is to realize that, to lowest order in $\epsilon$

\begin{equation}
\frac{C}{b_{2}}=\frac{3(\mu^{2}X Y+X' Y')^2+\mu^{2}(X' Y - X
Y')^2}{\mu^{2}(\mu^{2}X^2+\mu^{2}Y^2+X'^2+Y'^2)}.
\end{equation}

\noindent Knowing $C/b_{2}$ we use (\ref{Eq:xiMax1}) and (\ref{Eq:xiMax2}) to find the maximum value of $\xi$ for the particle, and hence the maximum values of $r_{1}$ and $r_{2}$. These maxima will eventually be reached (unless $\epsilon b_{2} = 0$ exactly), and so the particle will be guided indefinitely if these are smaller than the boundaries of the guide, and will be lost with certainty if they are larger.

We applied this algorithm to a large number of particles to find
the set that are transmitted by a guide in the
presence of a very small transverse coupling. We find that the
coupling reduces the acceptance of an infinitely long guide to
72\% of its acceptance in the absence of nonlinear forces. This is a general result that depends only on the coupling term being small enough
that our lowest order approximation is valid. In the absence of
coupling, the acceptance ellipses in the $(X,X')$ and $(Y,Y')$
phase spaces are uniformly dense - all particles that lie inside
both ellipses are transmitted. This ceases to be true when the
coupling is turned on. In fact, the sizes of the ellipses that
enclose the stable particles do not change, but these particles
are no longer distributed uniformly within them. For example, a
particle that is close to the boundary of the ellipse in $(X,X')$
may be driven out of the guide if it also happens to lie close to
the boundary in $(Y,Y')$, but it will be successfully transmitted
if it happens to lie close to the origin in $(Y,Y')$. Thus, the
phase space acceptance in one direction has a dense region around
the origin, becoming less dense towards the boundaries.

By eliminating the independent variable, $\sigma$, from (\ref{Eq:Diffxi}) and (\ref{Eq:Diffgamma}), we have obtained a great deal of insight into the coupled motion of the particles. However, we do not yet know the wavelength of the slow oscillations of amplitude and phase. To find this, we need to study the solutions of (\ref{Eq:Diffxi}) and (\ref{Eq:Diffgamma}) with $b_{1}=0$. There appears to be no straightforward analytical solution, so we solved the equations numerically for a set of initial conditions corresponding to the full range of the contours seen in figure \ref{Fig:Contours}, and for several values of the single parameter appearing in these equations, $\epsilon b_{2} E_{0}/(4\mu)$. In this way, we found the wavelength for the slow energy-exchanging oscillations (in units of $s$) to be

\begin{equation}
\lambda_{s} = \frac{4\mu\,g}{\epsilon b_{2} E_{0}}.\label{Eq:lambdaSlow}
\end{equation}

\noindent Here, $g$ is a number that depends only on the value of $C/b_{2}$, i.e. on the particular $(\xi,\gamma)$ contour that the particle travels along. For most of the contours, the value of $g$ is between 5 and 10. The shortest wavelengths ($g \simeq 5)$ are obtained for contours near the two maxima in figure \ref{Fig:Contours}, and near the edges of the plot where $\xi$ is 0 or 1. The longest wavelengths are found for values of $C/b_{2}$ near $1/4$ where the crossover from the closed contours to the open ones occurs. Taking an example where $\mu = 0.5$, $\epsilon b_{2}=0.1$, $E_{0}=r_{1}^{2}+r_{2}^{2}=1$ and $g=5$ we find $\lambda_{s}$ to be 100 unit cells of the guide.

\subsection{Larger coupling coefficient}

\begin{figure}
\centering
\includegraphics{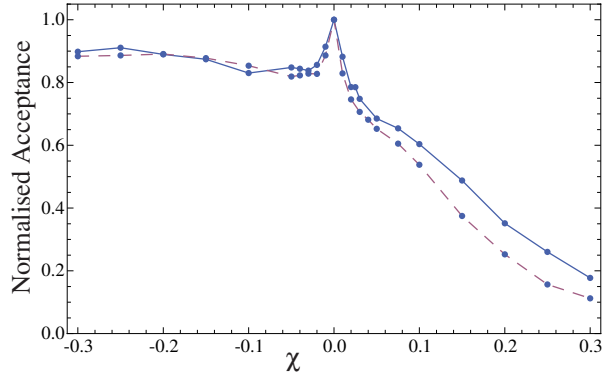}
\caption{Variation of the phase space acceptance with the value of
$\chi$. The guide has 100 lenses and has $\eta L =1$, $\eta S =
0$. The points joined by a solid line show the results of
numerical calculations with the micromotion included. The points
joined by a dashed line were calculated in the small micromotion
approximation. \label{Fig:CoupledAcceptance}}
\end{figure}

Having understood how the acceptance is reduced by a very small nonlinear coupling, we now consider larger values of the coefficient $\chi$. Applying the multiple scales method to higher orders in $\epsilon$ becomes quite awkward mathematically, so we turn to numerical methods. Then we have a choice: we could either solve the exact differential equations given by (\ref{Eq:EqnOfMotionComplete}), or we could use the small-micromotion approximation and solve (\ref{Eq:ReducedEqsOfMotion}) instead. In the former case, the step size needs to be chosen small relative to the micromotion wavelength, while for the latter it need only be small relative to the macromotion wavelength. For this reason, the approximate method is computationally faster by the ratio of the two wavelengths. Figure \ref{Fig:CoupledAcceptance} shows the results obtained by both methods for a gapless guide containing 100 lenses each having $\eta L =1$. The two methods give similar results, the greatest difference being for large positive values of $\chi$ where neglect of the micromotion underestimates the true acceptance.

In line with our discussion in section \ref{Sec:smallCoupling}, the acceptance drops on either side of the $\chi = 0$ position, due to the exchange of energy between transverse modes. This loss mechanism is fully established once $\chi$ is large enough that a cycle of the slow amplitude and phase oscillations is completed within the length of the guide. The width of the sharp peak depends inversely on this length, a longer guide resulting in a narrower central peak. Away from the central peak there is a striking asymmetry between positive and negative values of $\chi$. This happens because there are two effects that cause the acceptance to depend on $\chi$, one which reduces the acceptance for both positive and negative $\chi$, and another which reduces the acceptance on the positive side but increases it on the negative side. The first effect is the exchange of energy between the two transverse oscillations, already discussed above. Importantly, the wavelength of the amplitude and phase oscillations, $\lambda_{s}$, is a function of the particle's transverse energy, $E_{0}$, and its value of $C/b_{2}$. Those particles with $C/b_{2}$ near 1/4 will have longer energy-exchanging wavelengths than others, and so this loss mechanism continues to be active as $|\chi|$ increases. As a result the central peak in the figure has broad wings where there continues to be more particle loss with increasing $|\chi|$, reducing the acceptance on both sides of the central peak. The second effect is a direct result of the change in shape of the effective potential with the value of $\chi$. As $\chi$ increases, the nonlinear term makes the effective potential shallower along the diagonals of the guide, allowing an increasingly large fraction of the particles to escape along these directions. For negative values of $\chi$, the guide becomes deeper along the diagonals, allowing extra particles to be confined. It so happens that, for negative $\chi$, the two effects nearly cancel so that the acceptance is nearly a constant on the negative side.

\begin{figure}
\centering
\includegraphics[width=8cm]{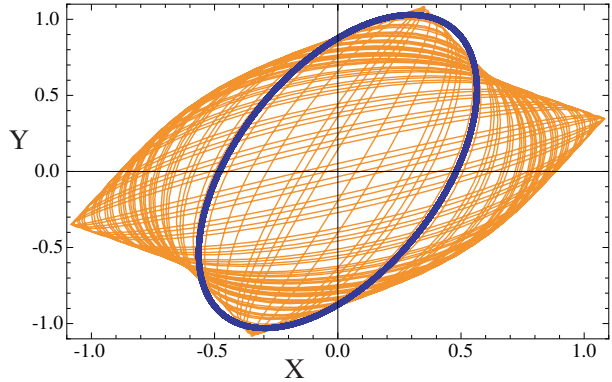}
\caption{Trajectory of a particle in the $X Y$-plane, for a
gapless guide with $\eta L= 1$. The initial conditions were
arbitrarily chosen. When $\chi = 0$ the trajectory is a closed
ellipse (thick line). When the transverse motions are coupled
(thin line, $\chi = -0.05$), the trajectory is more complicated
and the particle visits many more points in the plane.
\label{Fig:CoupledXYTrajectory}}
\end{figure}

It is also important to note that the coupling term leaves the shape of the effective potential unaltered along the principal axes. For all negative $\chi$, the potential depth is shallowest along these axes. In the presence of the coupling, particles tend to explore many different regions of the guide, and so for many particles, increasing the potential depth along the diagonals does not improve the confinement since these particles can still escape along the principal axes. Figure \ref{Fig:CoupledXYTrajectory} shows an example of the motion in the $X Y$-plane when $\eta L =1$, comparing $\chi = -0.05$ to $\chi = 0$. In the absence of coupling the particle moves in a closed ellipse. The coupling results in a far more complicated trajectory, with the particle visiting far more points in the plane.

Finally in this section, we consider the effect of the transverse coupling on the example guides that we examined earlier. In the case of the two-rod guide used in many previous experiments on focussing, guiding and decelerating, $\chi$ is always negative and therefore has little effect on the acceptance. This, along with its simplicity, is a great advantage of the two-rod geometry. The four rod guide with two-pole symmetry discussed in reference \cite{Bethlem(1)06} has $\chi = 0.19$ when the Stark shift is linear, and this reduces the acceptance to about 40\% of the value it takes in the absence of the coupling (for $\eta L = 1$ and $\eta S = 0$). When the Stark shift is quadratic $\chi = 0.055$ in this geometry and so the normalized acceptance is about 70\%. The four rod bent guide used for guiding molecules in reference \cite{Junglen04} has $\chi = 1.5$ when the Stark shift is linear. This very large value for $\chi$ is a great disadvantage of this geometry since it reduces the acceptance enormously. It is, of course, necessary to consider the higher order terms to get a complete picture in this case.

\section{Combined effect of both nonlinear terms}

\subsection{Small nonlinear coefficients}

\begin{figure}[t]
\centering
\includegraphics{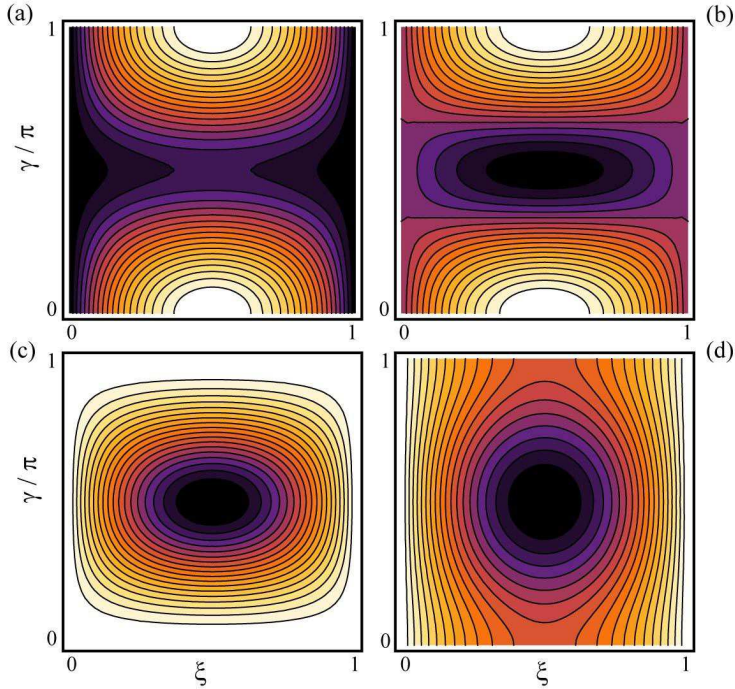}
\caption{Contour plot of (\ref{Eq:xigammacurves})
illustrating the slow oscillations in amplitude and phase. (a)
$b_{1}/b_{2}=0.25$, (b) $b_{1}/b_{2}=0.5$, (c) $b_{1}/b_{2}=1$,
(d) $b_{1}/b_{2}=2$. \label{Fig:ContoursBothTerms}}
\end{figure}

We return now to the general case where both $b_{1}$ and $b_{2}$
are nonzero. Figure \ref{Fig:ContoursBothTerms} shows contours of (\ref{Eq:xigammacurves}) for several different values of the
ratio $b_{1}/b_{2}$, along with colour maps to indicate the values, $C/b_{2}$, of the contours. As before, the fixed points in these plots
correspond to the coupled stationary modes. The ones at
$(\xi,\gamma)=(\case{1}{2},0),(\case{1}{2},\pi)$ are the diagonal in-phase and anti-phase
normal modes, while the fixed point at $(\case{1}{2},\case{\pi}{2})$ represents
the clockwise and anticlockwise circular modes. When $b_{1}/b_{2}$
is small, as in (a), the plot closely resembles the $b_{1}=0$ case
shown in figure \ref{Fig:Contours}. The main difference is that a greater fraction of the contours encircle the two maxima that correspond to the normal modes. Part (b) of the figure shows the contours for
$b_{1}/b_{2}=0.5$. The saddle point has been replaced by a minimum meaning that the circular modes, as well as the diagonal modes, are stable. All motion involves a
complete exchange of energy between one transverse direction and
the other. As $b_{1}/b_{2}$ is increased further, the area of the
plot containing contours centred on $(\case{1}{2},\case{\pi}{2})$ grows.
Eventually, once $b_{1}/b_{2}=1$ as in part (c) of the figure, this area includes all the
contours. With a ratio greater
than 1, e.g. $b_{1}/b_{2}=2$ as shown in (d), the diagonal modes
become unstable, and a larger fraction of the contours are associated with these modes.

It is clear from figure \ref{Fig:ContoursBothTerms} that the
coupled stationary modes can be stable or unstable, depending on
the ratio of $b_{1}$ to $b_{2}$. To find the stability condition,
we form the Hessian matrix of the left-hand-side of (\ref{Eq:xigammacurves}). A fixed point is a stable one if
the determinant of this matrix is positive. Using this procedure
it is straightforward to show that the normal modes are stable
provided $b_{1}/b_{2}<1$, and the circular modes are stable when
$b_{1}/b_{2}>1/3$.

We can also easily find the spatial frequencies of the coupled
stationary modes, following the same procedure as before. The
frequency in the $X$ direction is $\omega=\mu +
\epsilon\,d\theta_{1}/d\sigma$. With the help of (\ref{Eq:slowDiffs3}) we find the normal mode frequency to be

\begin{equation}
\omega_{{\rm NM}}=\mu\left(1-\frac{r_{1}^{2}}{8\mu^{2}}(3\epsilon b_{1}
+ 3\epsilon b_{2})\right),
\end{equation}

\noindent and the frequency of the circular mode to be

\begin{equation}
\omega_{{\rm CM}}=\mu\left(1-\frac{r_{1}^{2}}{8\mu^{2}}(3\epsilon b_{1}
+ \epsilon b_{2})\right).
\end{equation}

\noindent Comparing these with equations
(\ref{Eq:FrequencyExpansionCubic}) and
(\ref{Eq:FrequencyCoupled}), we see that, to this order in $\epsilon$, the change in frequency
resulting from the inclusion of both cubic and coupling terms
together is just the sum of the changes introduced by each term
alone.

\subsection{Larger nonlinear coefficients}

\begin{figure}[t]
\centering
\includegraphics{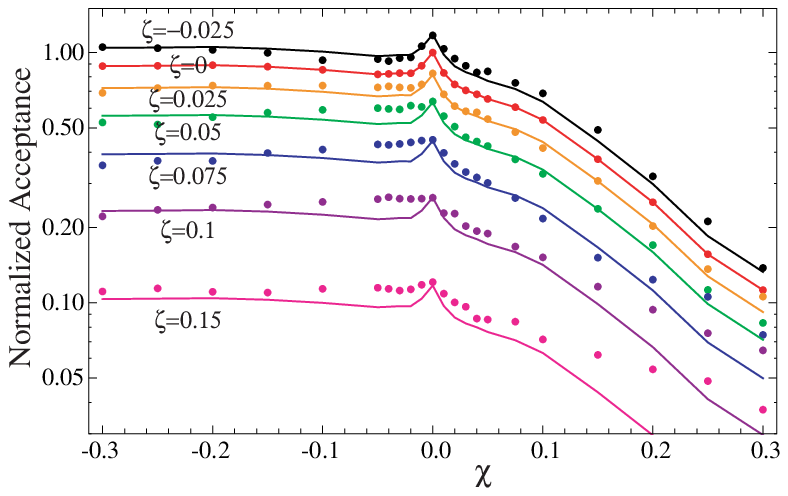}
\caption{Log-linear plot showing the 2D phase space acceptance of a guide with $\eta L = 1$,
$\eta S = 0$, plotted as a function of $\chi$ for various values
of $\zeta$. The acceptance has been normalized to the value
obtained when both $\chi$ and $\zeta$ are zero. The points are the
results of numerical simulations using the effective potential, (\ref{Eq:Veff}).
The solid lines show the product of the numerical acceptance
calculated with $\zeta = 0$ (as in figure \ref{Fig:CoupledAcceptance}) with the analytical result for
the cubic term, equations (\ref{Eq:cubicF1}) and
(\ref{Eq:cubicF2}).\label{Fig:BothTerms}}
\end{figure}

To find the 2D phase space acceptance over a wide range of $\zeta$ and $\chi$ we solved the equations of motion for a large number of particles using the effective potential given by (\ref{Eq:Veff}). The results for a 100-lens gapless guide with $\eta L=1$ are shown by the points in figure \ref{Fig:BothTerms}, for $-0.025 \le \zeta \le 0.15$ and $-0.3 \le \chi \le 0.3$. For all values of $\zeta$, the results are rather similar to the $\zeta = 0$ result shown in figure \ref{Fig:CoupledAcceptance} - there is a narrow spike centred at $\chi = 0$, a decline in acceptance for positive $\chi$ and a plateau for negative $\chi$. This similarity suggests a hypothesis, which we could write symbolically as $A(\zeta,\chi) = A_{0}F_{\zeta}(\zeta)F_{\chi}(\chi)$. In words, the hypothesis is that the total acceptance is the product of the one obtained from the linear theory, $A_{0}$, the factor $F_{\zeta}$ given by (\ref{Eq:cubicF1}) and (\ref{Eq:cubicF2}), which is independent of $\chi$ and accounts for the cubic term, and a third factor, $F_{\chi}$, that accounts for the coupling term independently of $\zeta$ and is plotted in figure \ref{Fig:CoupledAcceptance}. The lines in figure \ref{Fig:BothTerms} are plots of this product form, showing that for a large region of the parameter space this simple method agrees with the full numerical calculation to within a few percent. The largest discrepancies occur when both $\chi$ and $\zeta$ are large and positive. Here, the numerical calculations show that the acceptance is considerably larger than the product form suggests, e.g. 70\% larger when $\zeta=0.15$, $\chi=0.3$. Away from this region, the next largest discrepancies occur when $\chi$ is small and negative. The numerical calculations show that in this region the acceptance tends to be larger than the product form predicts, by up to 20\%.

We are, at last, ready to put everything together to find the best
way of building alternating gradient guides, decelerators and
traps. We first found that, in the linear theory, the 2D
phase-space acceptance is proportional to $\eta^{2}$, and so to
the value of $|a_{3}|$. However, making $a_{3}$ as large as
possible may not be the best strategy, because of the detrimental
effects of positive $\zeta$ and $\chi$ which are related to one
another through $a_{3}$: $\chi+3\zeta = 2n|a_{3}|$. If we make
$a_{3}$ too large, the increased acceptance suggested by the
linear theory will not be realizable due to the resulting large
value of $\chi$. Decreasing $\chi$ by increasing $\zeta$ does not
necessarily help, since large positive $\zeta$ also reduces
acceptance. In the effective potential picture, increasing $a_{3}$
increases the depth of the potential, but increasing either
$\zeta$ or $\chi$ decreases the depth. We are only free to choose
two of the three parameters, $a_{3}$, $\zeta$, $\chi$. What values
should we choose to optimize the acceptance? We will look at some
examples for particular values of the operating parameters, $\eta
L$ and $\eta S$.

\begin{figure}
\centering
\includegraphics{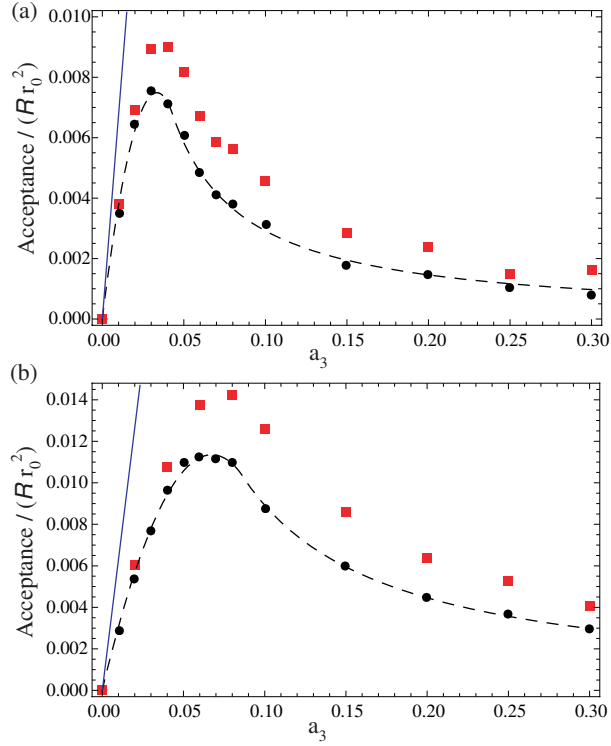}
\caption{2D acceptance versus $a_{3}$ for an AG guide with a
two-rod geometry, in the case of a linear Stark shift. (a) Guide with no gaps: $\eta L =1$, $\eta S = 0$. (b) Guide with gaps and lenses of equal length: $\eta L =\eta
S = 0.7$. Dashed lines are analytical results. Round points are
results of numerical simulations based on the effective potential.
Square points are results of full numerical simulations. The solid line is the result obtained in the absence of nonlinearities.
\label{Fig:2RodVa3}}
\end{figure}

We start with the two-rod geometry. This is the easiest
case to consider since there is only one free parameter, $a_{3}$,
determined by the the ratio of the rod radius, $R$, to the gap
between the rods, $2r_0$: $a_{3}=(r_{0}/R)/(2+r_{0}/R)$ and
$a_{5}=a_{3}^{2}$. The two nonlinear coupling coefficients are
therefore $\zeta=2|a_{3}|$ and $\chi=-4|a_{3}|$ when the Stark
shift is linear, and $\zeta =3|a_{3}|$, $\chi = -5|a_{3}|$ when
the Stark shift is quadratic. The calculation is hugely simplified
by $\chi$ being negative because we know that the acceptance is
insensitive to $\chi$ when it is negative, and we have an
approximate analytical expression to describe the effect of the
cubic nonlinearity. Our question can therefore be answered
analytically with good accuracy.

Figure \ref{Fig:2RodVa3}(a) shows the transverse acceptance versus
$a_{3}$ for a two-rod gapless guide with $\eta L = 1$, in the case of a linear Stark shift. The unit of the acceptance is ${\cal R}r_{0}^{2}$ and is independent of $a_{3}$. See equation (\ref{Eq:etaSqLin}) for the definition of ${\cal R}$. The solid line in the figure is the analytical result obtained in the absence of nonlinearities, (\ref{Eq:exampleAcceptance}). The dashed line in the figure is also a completely analytical result - we take the result given by the solid line, multiply by the correction factor given by (\ref{Eq:cubicF1}) and
(\ref{Eq:cubicF2}) to account for the cubic nonlinearity, and
finally multiply by 0.9 to account for the negative value of
$\chi$, as suggested by figure \ref{Fig:CoupledAcceptance}. As
$a_{3}$ is increased from zero the acceptance first increases
linearly, just as expected from the theory when the nonlinearities
are neglected. However, as $a_{3}$ increases further, the growing
cubic nonlinearity halts and then reverses the ascent of the
acceptance, which reaches its maximum value near $a_{3}=0.03$. We
note that the rod radius needs to be $16r_{0}$ to obtain this
value of $a_{3}$. The maximum acceptance is approximately $0.008
{\cal R}r_{0}^{2}$. We will examine this value in the context of
some typical experiments at the end of the paper. The round points
in the same figure show the results of numerical calculations of
the transverse acceptance using the effective potential approach.
Our analytical result agrees very well with these numerical
calculations. The square points show the results obtained from a
full numerical calculation, including the micromotion. As in
previous examples, the inclusion of the micromotion increases the
acceptance, but the overall shape of the plot remains unchanged,
and the maximum of the acceptance still occurs close to the same
value of $a_{3}$. We have shown the result for a gapless guide
operated at $\eta L = 1$ which is close to the optimal lens length
(see figure \ref{Fig:LinAcc}). Results for other cases are just as
easily obtained using the same procedure. In general, the optimal
value of $a_{3}$ increases as $\eta L$ is increased towards the edge of stability. Figure \ref{Fig:2RodVa3}(b) provides the same
information for a guide with gap lengths equal to the lens
lengths. This is the situation typically found in AG stark
decelerators. Again, we chose the parameters close to the optimal
ones in the linear theory, specifically $\eta L = \eta S = 0.7$.
The figure shows exactly the same trends as for the gapless case,
but the acceptance is now maximized when $a_{3} \simeq 0.07-0.08$,
requiring rods of radii 6-7$r_{0}$. Again, the numerical
simulations based on the effective potential agree very well with
the analytical result, while the full numerical simulations show
that the true acceptance is a little larger.

\begin{figure}
\centering
\includegraphics{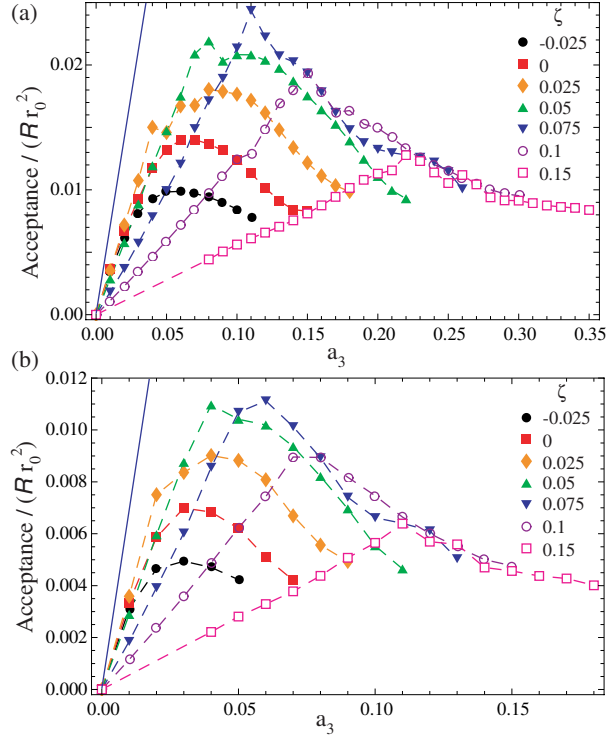}
\caption{2D acceptance as a function of $a_{3}$, when $\eta L =
1$, $\eta S = 0$, calculated for several values of $\zeta$. The
dashed lines simply join the points to improve the clarity of the
figure. The solid line gives the result in the absence of nonlinearities. (a) Linear Stark shift, (b) Quadratic Stark shift.
\label{Fig:GeneralVa3}}
\end{figure}

Returning to the more general case, we will take $a_{3}$ and
$\zeta$ to be the two free parameters, $\chi$ being fixed once
these two are chosen. We will concentrate on the example of a
gapless guide operated with $\eta L = 1$. Then, there are no new
calculations to do since all the information we need is already
given in figure \ref{Fig:BothTerms}. For each value of $a_{3}$ and
$\zeta$, we calculate the acceptance in the linear case, which is
just linear in $a_{3}$, and then read off the multiplication
factor from figure \ref{Fig:BothTerms}.

We present the results, for both linear and quadratic Stark
shifts, in figure \ref{Fig:GeneralVa3}. Let us concentrate first on
the linear case, part (a) of the figure. We know from figure \ref{Fig:Cubic} that negative values of $\zeta$ increase the
acceptance above the value given by the linear theory, and it is
tempting to conclude that we should design the guide with negative
$\zeta$. However, negative $\zeta$ implies large positive values
for $\chi$ which greatly reduces the acceptance. As a result, the
optimal acceptance is small and occurs for small values of
$a_{3}$, as illustrated by the results for $\zeta = -0.025$ shown
in figure \ref{Fig:GeneralVa3}(a). The figure shows that increasing
the value of $\zeta$ increases the maximum acceptance, and also
the optimal value of $a_{3}$. This is because the beneficial
effect of a smaller $\chi$ outweighs the detrimental effect of a
larger $\zeta$. The acceptance reaches its maximum value of $0.025 {\cal R}r_{0}^{2}$ when $\zeta \simeq 0.075$ and $a_{3} \simeq 0.11$. At this point, $\chi$ is very
close to zero. For values of $a_{3}$ below the optimum, $\chi$ is
negative and so has very little influence on the acceptance, which
consequently increases rather linearly with $a_{3}$. As soon as
$\chi$ becomes positive further increases in $a_{3}$ only reduce
the acceptance because of the associated increases in $\chi$. When
$\zeta > 0.075$, the same trends are seen - the acceptance
increases linearly with $a_{3}$ up to the point where $\chi$ is
zero, and then decreases. Because increasing $\zeta$ reduces the
acceptance, the slope of this linear rise is smaller than before,
the acceptance is maximized at higher values of $a_{3}$, and this
maximum value is reduced. For comparison, the result obtained in the absence of nonlinearities is shown by the solid line in the figure.

The results for a quadratic Stark shift are shown in figure \ref{Fig:GeneralVa3}(b). The figure shows exactly the same
characteristics as for the linear Stark shift, but with all
features occurring at lower values of $a_{3}$. The acceptance
obtains its maximum value when $\zeta \simeq 0.075$ and $a_{3}
\simeq 0.06$. As before, $\chi \simeq 0$ at this point, and this
is also true of the turning points for higher values of $\zeta$.
The maximum acceptance is $0.011 {\cal R} r_{0}^{2}$, about half
the value obtained for the linear Stark shift.

The recipe for designing an AG guide should now be clear - find an
electrode structure that produces a small positive value of
$\zeta$ (close to 0.075), and a $\chi$ very close to zero, and
choose lens lengths such that $\eta L$ is close to 1. If there is
a need for gaps between the lenses, such as in an AG decelerator,
it is best to keep the gaps short, in which case the above
conclusions are not much altered.

\section{Conclusions}

In this paper we have shown how nonlinear forces alter the
trajectories of neutral particles in an AG guide, and have shown
how to calculate the phase space acceptance of guides with these
nonlinear forces included. We have found how to design the guide
so that the acceptance is optimized. To conclude, we apply our
results to some example experiments, revisiting those discussed in
Sec. \ref{Sec:LinearPSA}.

Consider a beam of CaF molecules entering an AG guide with a
forward speed of 350\,m/s. The guide has $r_{0} = 1$\,mm, $E_{0} =
100$\,kV/cm, $\eta L = 1$ and $\eta S = 0$, and the electrode
geometry is designed so that $a_{3}$ and $\zeta$ have optimal
values, as in figure \ref{Fig:GeneralVa3}. The Stark shift will be
linear in this case, and the ratio of the Stark shift at the
centre of the guide to the forward kinetic energy is ${\cal R} =
1/88$. The transverse acceptance will be approximately $0.025
{\cal R} r_{0}^{2} = (17\,{\rm mm\,mrad})^{2}$. The equivalent
trap depth of this guide is approximately 30\,mK. The available
acceptance is well matched to the phase space area occupied by the
beam in a typical supersonic beam experiment where the molecules
pass through a small skimmer before entering the guide. We can
therefore expect the guide to transmit a large fraction of the
total beam. For a second example we consider Li atoms launched
with a speed of 10\,m/s from a MOT, into an AG guide. The guide
has the same parameters as above, except that $a_{3}$ is halved so
that the acceptance is optimized for the quadratic Stark shift. In
this case ${\cal R}=0.47$, and the transverse acceptance is
approximately $0.01{\cal R} r_{0}^{2} = (70\,{\rm mm\,mrad})^{2}$.
The equivalent trap depth of the guide is approximately
50\,$\mu$K, similar to the temperature of the atoms in the MOT.
Again, we can expect the guide to transport a large fraction of
the atoms. Unless the length of the guide is shorter than the
macromotion wavelength, in which case the transmission will be
larger than the above estimates, the acceptance does not depend on
the length. The guide can be made arbitrarily long,
delivering atoms and molecules to an experiment far removed from
the source. Our examples demonstrate that a carefully designed AG
guide has an acceptance that is adequate for many common
experimental situations, making it a useful tool for cold atom and
molecule physics.

\ack

We are indebted to Jony Hudson and Ben Sauer for their help in improving this paper. We thank Rick Bethlem, Jochen K\"{u}pper and Gerard Meijer for many valuable discussions on the subject of alternating gradient focussing.  We are grateful to the Royal Society for supporting the authors, and to the STFC and the EPSRC for their financial support.

\end{document}